\documentclass[aps,prx,amsmath,nofootinbib,longbibliography,amssymb,superscriptaddress,twocolumn,10pt]{revtex4-2}
\usepackage[english]{babel}
\usepackage{graphicx}
\usepackage{ulem}
\usepackage{soul}
\usepackage{float} 
\usepackage[svgnames]{xcolor} 
\usepackage{comment}
\usepackage[colorlinks=false]{hyperref}
\usepackage{verbatim}
\usepackage{esint}
\usepackage{marginnote}
\usepackage{subfigure}
\newcommand{\beq} {\begin{equation}}
\newcommand{\eeq} {\end{equation}}
\newcommand{\bea} {\begin{eqnarray}}
\newcommand{\eea} {\end{eqnarray}}
\newcommand{\be} {\begin{equation}}
\newcommand{\ee} {\end{equation}}

\begin{document}
\title{Pairing at a single Van Hove point}
\author{Risto Ojajärvi}
\affiliation{Institut für Theorie der Kondensierten Materie, Karlsruher Institut
für Technologie; 76131 Karlsruhe, Germany}
\author{Andrey V. Chubukov}
\affiliation{W. I. Fine Theoretical Physics Institute, University of Minnesota,
Minneapolis, MN 55455, USA}
\author{Yueh-Chen Lee}
\affiliation{W. I. Fine Theoretical Physics Institute, University of Minnesota,
Minneapolis, MN 55455, USA}
\author{Markus Garst}
\affiliation{Institut für Theoretische Festkörperphysik, Karlsruher Institut für
Technologie; 76131 Karlsruhe, Germany}
\affiliation{Institut für QuantenMaterialien und Technologien, Karlsruher Institut
für Technologie; 76131 Karlsruhe, Germany}
\author{Jörg Schmalian}
\affiliation{Institut für Theorie der Kondensierten Materie, Karlsruher Institut
für Technologie; 76131 Karlsruhe, Germany}
\affiliation{Institut für QuantenMaterialien und Technologien, Karlsruher Institut
für Technologie; 76131 Karlsruhe, Germany}
\begin{abstract}
We show that an interacting electronic system with a single ordinary or extended Van Hove point, which crosses the Fermi energy, is unstable against triplet superconductivity. The pairing mechanism is unconventional. There is no Cooper instability. Instead, pairing is due to the divergence of the density of states at a Van Hove point, leading to a superconducting quantum critical point at a finite detuning from the Van Hove point. The transition temperature is universally determined by the exponent governing the divergence of the density of states. Enhancing this exponent drastically increases $T_c$. The Cooper pair wave function has  a non-monotonic momentum dependence with a steep slope near the gap nodes. In the absence of spin--orbit coupling, pairing fluctuations suppress a $2e$ spin-triplet state, but allow pairs of triplets to condense into a charge-$4e$ singlet state at a temperature of similar order as our result.
\end{abstract}
\maketitle

\section{Introduction}

The ability to create and manipulate two-dimensional
(2D)
 or strongly anisotropic 3D
 electronic materials led to an increased interest in the theory of
systems
 in which
 the Fermi energy is at or near a Van Hove
 (VH)
  singularity
of the electronic density of states
 ~\cite{VanHove1953}.
 Examples are doped graphene\cite{Nandkishore2012,Kiesel2012,Wang2012,BlackSchaffer2014},
a wide range of moiré materials\cite{Shtyk2017,Yuan2019,Isobe2019,Yuan2020,Chandrasekaran2020,Classen2020,Cano2021},
metallic Kagome systems\cite{Wang2013,Hu2022,Kang2022},
as well as the
 ruthenate oxides  Sr$_{3}$Ru$_{2}$O$_{7}$
 in an external magnetic field\cite{Efremov2019} and
 Sr$_{2}$RuO$_{4}$ under uni-axial compressive strain~\cite{Hicks2014,Barber2018,Li2022,Stangier2022}.
 Studies of these materials extended
  earlier
   theoretical
    analysis of
 VH
 singularities
 in cuprate
  and other
   superconductors
    \cite{Dzyaloshinskii1987,Dzyaloshinskii1988,
    Virosztek1990,Newns1995,
  Hlubina1996,Furukawa1998,Alvarez1998,Honerkamp2001,Irkhin2001,
   LeHur2009,Husemann2009,Dzyaloshinskii1996,Menashe1999}.

\begin{figure}
    \centering
    \includegraphics[width=\columnwidth]{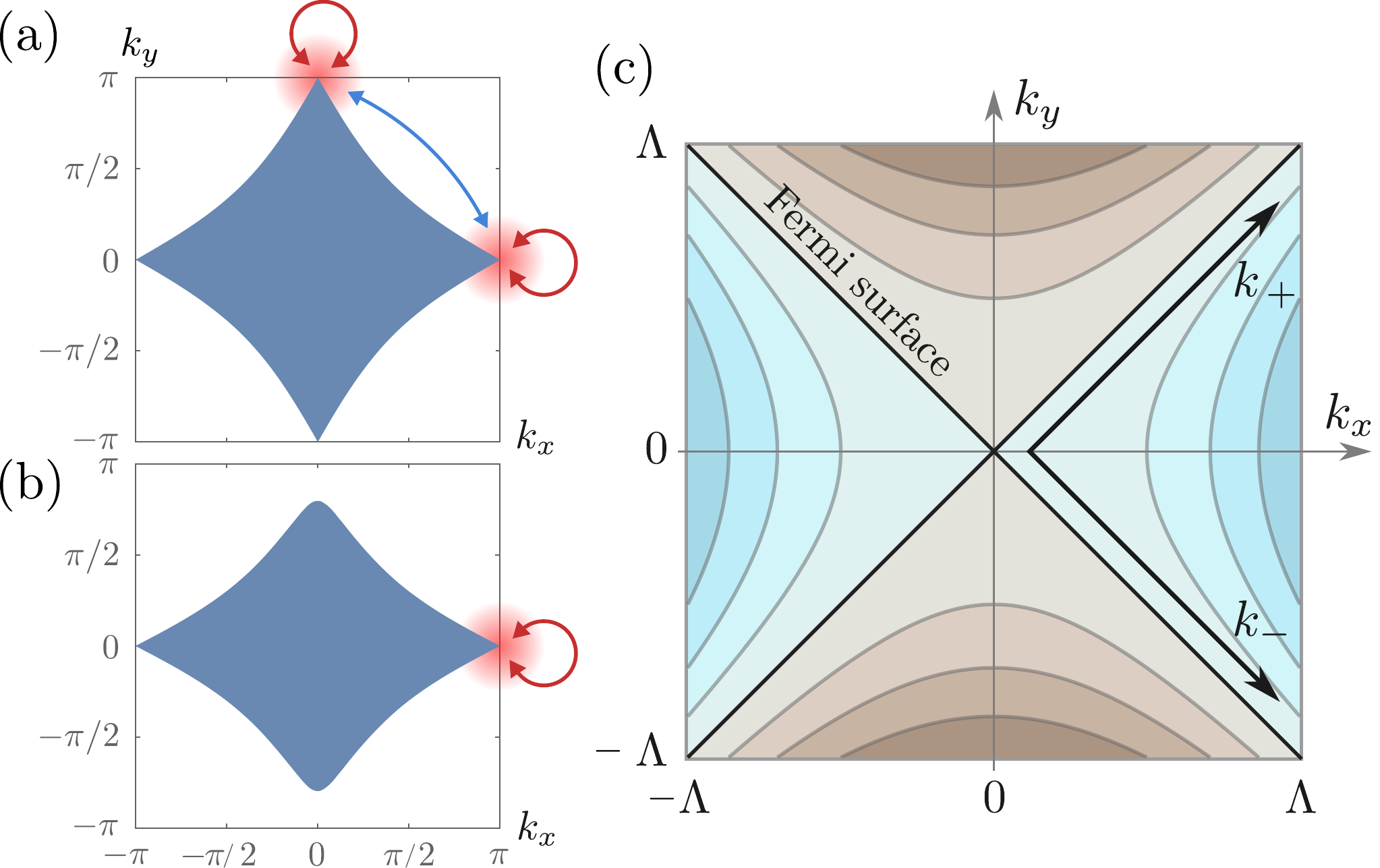}
    \caption{Fermi surface of a system with (a) two Van Hove points, and (b) a single Van Hove point at the Fermi energy. The shaded area shows the occupied states, and arrows illustrate the dominant interactions.
    (c) Iso-energy contours near the Van Hove point described by Eq.~\eqref{eq:VHdisp}. Blue (brown) color indicates the occupied  (unoccupied) states. 
     The Fermi surface for the extended Van Hove point  of Eq.~\eqref{eq:extVH} is the same, while the iso-energy contours will be different, with flatter bands near the Fermi energy. The coordinates
     $k_{\pm} = (k_x \pm k_y)/\sqrt{2}$
     are extensively used throughout the text, where we consider gap functions that depend on only one of these coordinates, i.e. $\Delta(k_+)$ or $\Delta(k_{-})$.  }
    \label{fig:fermisurface}
\end{figure}

  In the analysis of the impact of VH singularities,
 two cases
 should be distinguished: 
 In the first case, illustrated in Fig.~\ref{fig:fermisurface}(a), there are several
 symmetry-related VH
 points
that simultaneously cross the Fermi energy.
In this situation,
scattering events with
large transferred momentum that connect different VH
 points
are crucial. These processes often lead to density-wave instabilities,
   which trigger superconductivity nearby in the phase diagram via some version of the Kohn-Luttinger mechanism~\cite{Kohn1965,Chubukov1993,Shankar1994,Arovas2022}.
In the second case,
 shown in Fig.~\ref{fig:fermisurface}(b),
there is
a single VH
point
at the Fermi level.
Such a situation
 occurs in systems with
low symmetry, including strained materials, where the application
of large uni-axial stress reduces the number of allowed symmetry operations. A prominent example
is Sr$_{2}$RuO$_{4}$, where stress along the Ru-O-Ru bond direction moves
one of the VH points of the tetragonal system to the Fermi energy,
while the other is pushed
away from it~\cite{Hicks2014,Barber2018,Li2022,Stangier2022}.

In addition to ordinary VH points, where the density of states in 2D diverges like a logarithm, extended VH points have recently been discussed extensively\cite{Shtyk2017,Yuan2019,Isobe2019,Yuan2020,Chandrasekaran2020,Classen2020}. In this  case 
the density of states diverges by a power law
  due a saddle-point that is less dispersive than the ordinary quadratic one.
 In Refs.~\cite{Yuan2019,Chandrasekaran2020} a classification of such extended singularities was given and it was shown that some extended VH points can be reached by solely varying a single parameter in the Hamiltonian.

For a single,
 ordinary or extended
 VH point,
a
 Stoner-type 
 analysis suggests a ferromagnetic instability at arbitrarily small interaction, due to the
  divergent  density of states~\cite{Irkhin2001,
  Cano2021}.
  However,
  no instability was
 detected
  in renormalization group
(RG)
  studies~
\cite{Isobe2019,Classen2020}.
 The authors of Ref.~\cite{Isobe2019} argued, based on their RG study, that
the ground state of a system with a single extended VH point
  is a particular non-Fermi liquid, dubbed a supermetal, in which  the quasi-particle weight vanishes
   by
  a power-law as one approaches low energies.

In this paper,
  we consider the behavior of a
 system of fermions with repulsive interaction and a
single ordinary
(extended)
VH
point in 2D with
logarithmic (power-law) divergence of the density of states.
 We
   first present
  arguments
  in favor of
  the absence of a ferromagnetic Stoner instability and then
  demonstrate that
 the divergence in the density of states gives rise to odd-parity spin triplet superconductivity.
  To obtain such an instability we had to go beyond
  the
   conventional 
     one-loop 
     RG treatment,
    which accounts for the leading logarithms, 
    but neglects the subleading ones. 
 For an ordinary VH point, we
  obtain for the superconducting transition temperature 
\begin{equation}
T_{c} = T_0 \exp\left(-\frac{1+ \mu g} {\gamma g}\right).
\label{eq:TcVH}
\end{equation}
 Here,
 $g$ is the dimensionless coupling constant to be defined below,
  $\gamma$
   and $\mu$ are
   of order one,
  and
  $T_0=\frac{\Lambda^{2}}{2m}$,
   where
 $m$
  is the curvature of the quadratic saddle-point
dispersion around a VH point and $\Lambda$ is the upper momentum cutoff of the theory.
  For an extended VH point we obtain
 \begin{equation}
      T_{{\rm c}}\sim T_{0}g^{\frac{1+\epsilon}{2\epsilon}},
     \label{eq:TceVH}
 \end{equation}
 where $\epsilon$ is the exponent
   that
   determines
   the
divergence of the density of states,
  $\rho (\omega) \propto |\omega|^{-\epsilon}$.
$T_c$  of  an extended Van Hove point is no longer exponentially small
 and 
 gets strongly enhanced when
$\epsilon$  increases.  Furthermore, we show that this $T_c$ is cutoff-independent as the dependence on $\Lambda$ cancels out between $T_0$ and $g$.
 Our result indicates that
the supermetal of Ref.~\cite{Isobe2019}
 describes
the normal state over
 some temperature and energy range, but
  at lowest temperatures or energies
  the system eventually
   becomes unstable against triplet superconductivity.

Solving the gap equation for $T \lesssim T_c$, we
  find that the
pairing state
 is highly non-local with an unusual momentum
dependence of the Cooper-pair wave function.
 It changes sign under $\boldsymbol{k}\rightarrow-\boldsymbol{k}$
 as required for odd-parity triplet pairing.
However, the momentum regime,
 where 
 the gap is linear in $\boldsymbol{k}$, 
 turns out
to be
 extremely small,
 of order $\delta k\sim2mT_{c}/\Lambda$ in the case of an ordinary VH point.
Nodal excitations should therefore be hardly visible in thermodynamic
measurements such as the specific heat, the Knight shift or the superfluid
stiffness.

  We also analyze
  the role of
 pairing
 fluctuations. The results
 Eqs.~\eqref{eq:TcVH}  and \eqref{eq:TceVH}  are mean-field transition temperatures.
 For a spin-singlet superconductor, the actual
 transition at $T_\text{BKT} \leq T_c$ is of
   Berezinskii--Kosterlitz--Thouless (BKT)
     type \cite{Kosterlitz1973,Kosterlitz1974}  into a
    charge-$2e$ state with algebraic order (power-law decay of superconducting correlations).
  For a 2D spin-triplet state,  a charge-$2e$ order
   survives in the presence of
   spin-orbit
interaction. In its absence a
spin-triplet order   is additionally suppressed by
fluctuations in the spin sector of the
 superconducting
 order parameter~\cite{Korshunov1985,Schmalian2021}.  However,
  there exists a BKT transition into to a charge-$4e$ state,
 in which
  two triplets bind into a singlet.
 In either case,
 the onset temperature for the
algebraic superconductivity is
  comparable to the
  mean-field
  $T_{c}$, given in Eqs.~\eqref{eq:TcVH} and \eqref{eq:TceVH}.
 
\section{The Model}
 We consider a system of interacting electrons with
  dispersion $\varepsilon_{\boldsymbol{k}}$ and
 Hubbard repulsion $U$:
\begin{equation}
H=\sum_{\boldsymbol{k}\alpha}\varepsilon_{\boldsymbol{k}}\psi_{\boldsymbol{k}\alpha}^{\dagger}
\psi_{\boldsymbol{k}\alpha}+\frac{U}{N}\sum_{\boldsymbol{k}\boldsymbol{k'}\boldsymbol{q}\alpha\beta}
\psi_{\boldsymbol{k}\alpha}^{\dagger}\psi_{\boldsymbol{k}'\beta}^{\dagger}\psi_{\boldsymbol{k}'-
\boldsymbol{q}\beta}\psi_{\boldsymbol{k}+\boldsymbol{q}\alpha},
\label{eq:H}
\end{equation}
where $\psi_{\boldsymbol{k}\alpha}$ annihilates an electron with
momentum $\boldsymbol{k}$ and spin $\alpha$. We measure the momenta relative to the VH point,  assumed to be time-reversal symmetric, and focus on the case where $\varepsilon_{\boldsymbol{k}=\boldsymbol{0}} =0$, i.e.,
 the VH point is right at the Fermi level. In the last section, we comment on the behavior upon tuning the Fermi energy away from the VH point.

For an ordinary VH point
 the electronic dispersion is
\begin{equation}
\varepsilon_{\boldsymbol{k}}=\frac{1}{2m}\left(k_{x}^{2}-k_{y}^{2}\right).\label{eq:VHdisp}
\end{equation}
An anisotropy between $k_x$ and $k_y$,
  expected
 for a VH point located away from the center of the Brillouin zone,
can be eliminated by an appropriate re-scaling of momenta.
The
quadratic
 dispersion in  Eq.~\eqref{eq:VHdisp} gives rise to a
   logarithmically
 diverging density of states $\rho\left(\omega\right)\sim m \log\left(\frac{\Lambda^{2}}{m\left|\omega\right|}\right)$,
 where
 $\Lambda$ is the momentum cutoff
--
 the highest momentum, up to which Eq.~\eqref{eq:VHdisp} is valid.

For an extended VH point, we follow earlier works~\cite{Furukawa1998,Honerkamp2001,LeHur2009,Husemann2009}
 and consider the dispersion in the form
\begin{equation}
\varepsilon_{\boldsymbol{k}}=A\left(\left|k_{x}\right|^{n}-\left|k_{y}\right|^{n}\right),\label{eq:extVH}
\end{equation}
where $n\geq2$. Now the density of states diverges
 by a power-law
$\rho\left(\omega\right)
\sim
 A^{\epsilon-1}\left|\omega\right|^{-\epsilon}$
with $\epsilon=1-2/n\geq0$.
 In this paper, we focus on the dispersion Eq.~\eqref{eq:extVH}, but note
that other
extended
 VH singularities are possible, e.g.,
with different
 powers for the two components of $\boldsymbol{k}$
 \cite{Shtyk2017,Yuan2019,Isobe2019,Yuan2020,Chandrasekaran2020,Classen2020}.

\section{A potential Stoner instability }
We begin with the discussion of a potential instability towards ferromagnetism.
 At
 a
 first glance, ferromagnetism near a single VH point is
a natural option~\cite{Irkhin2001},
  as this is  a $\boldsymbol{q}=0$ instability, and it develops for a repulsive
 interaction between fermions.
  Within  the random phase approximation (RPA),
 the instability  occurs when the dimensionless interaction -- the product of $U$ and the density of states -- reaches a certain finite value. In standard situations, this requires that the interaction strength exceeds a threshold value.  For the VH case, however, the density of states is divergent, and
  within RPA  the
    Stoner condition is satisfied
     already for an arbitrarily weak interaction.

\begin{figure}
    \centering
    \includegraphics[width=\columnwidth]{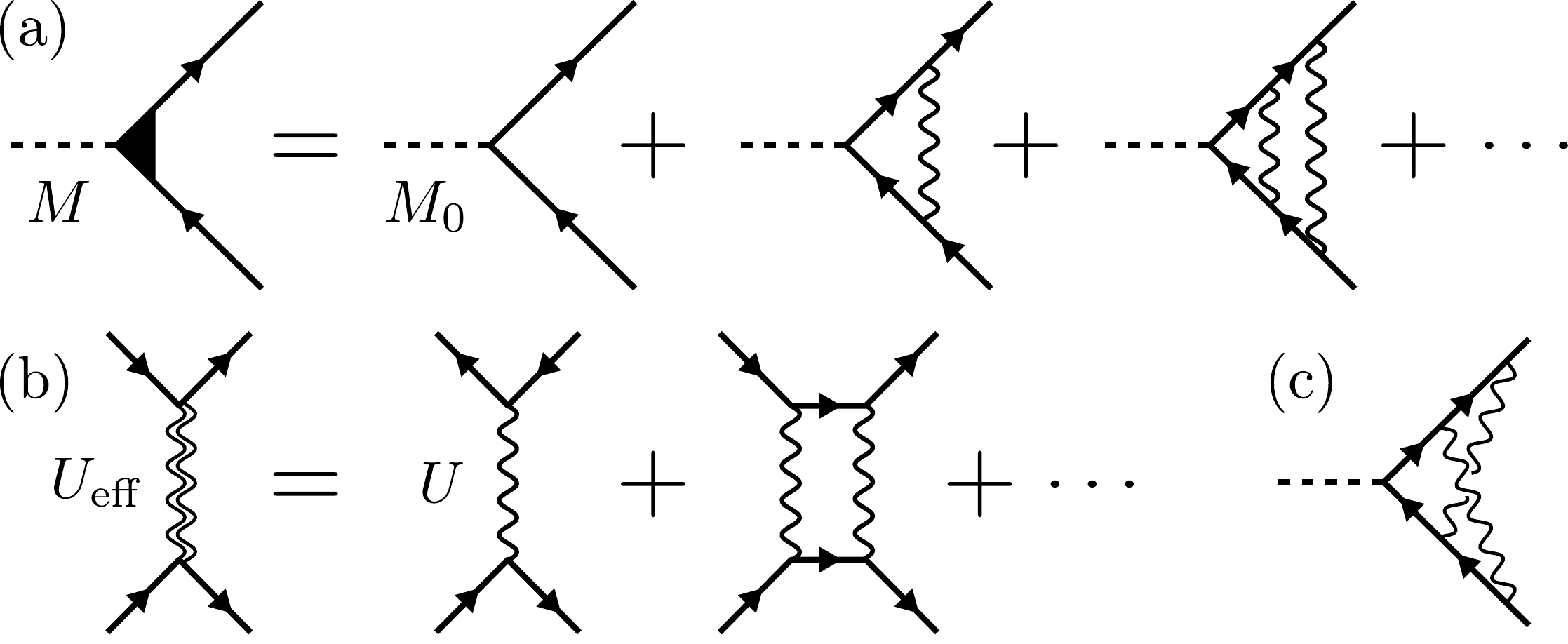}
    \caption{
     Diagrammatic representation of the fully-dressed order parameter $M$ in terms of the infinitesimally small bare one $M_0$. (a) Ladder series in the particle-hole channel [Eq.~\eqref{n_1}]. (b) Series of particle-particle diagrams [Eq.~\eqref{eq:eos}] which renormalize the interaction. (c) A 
     diagram, generated by 
      inserting a particle-particle renormalization into 
      the particle-hole ladder series.
       At 
        a Stoner transition, the diagrammatic series from $M/M_0$ must sum to
     infinity; see text for further details.}
    \label{fig:Stoner}
\end{figure}

  Whether or not a  Stoner instability
     develops beyond RPA
   can be detected by computing the static and uniform magnetic susceptibility $\chi$, i.e. the limit $\boldsymbol{q}\to 0$ of the static susceptibility $\chi (\boldsymbol{q},\omega=0)$.  This susceptibility can be obtained by either introducing an infinitesimal magnetic field and computing the magnetization or by introducing an infinitesimal bare ferromagnetic order parameter, dressing it by interactions,  and computing the ratio of the fully-dressed and the bare order parameters. In the diagrammatic analysis, the second approach is easier to implement. The bare order parameter  $M_0$ is represented as a two-particle vertex, and the  dressed one $M$ is obtained by renormalizing this vertex  by interactions. This is illustrated in Fig.~\ref{fig:Stoner}(a).

For definiteness, we consider a single ordinary VH point.
 The particle-hole polarization bubble at a VH point 
 is
   $\Pi_{{\rm ph}} = \tfrac{m}{2\pi^2} \log\frac{T_{0}}{T}$
 with $T_{0}$ given below Eq.~\eqref{eq:TcVH}.
 Summing up ladder (RPA) series of particle-hole renormalizations of $M_0$ one obtains
 \begin{equation}
 M = M_0 \left( 1 + U\Pi_{{\rm ph}} + \left(U\Pi_{{\rm ph}}\right)^2 + ...\right) = \frac{M_0}{1-U\Pi_{{\rm ph}}}.
 \label{n_1}
 \end{equation}
 The susceptibility $M/M_0$ diverges at $U\Pi_{{\rm ph}} =1$, i.e., at $T = T_{\rm FM}$ satisfying
  $\lambda_0 \log\frac{T_{0}}{T_{\rm FM}} =1$, where
  \begin{equation}
\lambda_{0}=\frac{mU}{2\pi^{2}}.
\end{equation}
The  ferromagnetic transition temperature $T_{\rm FM}$ is finite no matter how small $U$ is.

 This analysis, however, does not hold beyond RPA, once we include crossed diagrams (Fig.~\ref{fig:Stoner}(c)), which represent insertions of particle-particle renormalizations into the particle-hole channel.
  At zero total momentum, a
  particle-particle polarization bubble
  $\Pi_{{\rm pp}} = \tfrac{m}{2\pi^2} \log^{2}\frac{T_{0}}{T}$
 diverges as $\log^2$
  due to
  an additional Cooper logarithm.
  If we use this expression and add the  ladder series of particle-particle renormalizations to each term
  in Eq.~(\ref{n_1}), we effectively replace $U$ by
 \begin{eqnarray}
U_{\rm eff} &=&  U \left(1-U\Pi_{{\rm pp}}+U^{2}\Pi_{{\rm pp}}^{2}\cdots\right)\label{eq:eos} \nonumber \\
 & = &
  \frac{U}{1+\lambda_{0}\log^{2}\frac{T_{0}}{T}}.
\end{eqnarray}
 Substituting $U_{\rm eff}$ instead of $U$ into Eq. (\ref{n_1}), we find
 that the Stoner condition becomes
 \begin{equation}
\frac{\lambda_{0}\log\frac{T_{0}}{T_{\rm FM}}}{1+\lambda_{0}\log^{2}\frac{T_{0}}{T_{\rm FM}}}=1.\label{eq:Stoner_naive}
\end{equation}
  We show in Appendix A that
  the same result is obtained by 
  using a renormalization
group analysis with flow parameter $l=\log^{2}\frac{\Lambda}{k}$.
There is no solution of Eq.~\eqref{eq:Stoner_naive} at small $\lambda_0$
  because the  suppression of $\lambda_0$ by fluctuations in the particle-particle channel is stronger than the enhancement of $\lambda_0$ in the particle-hole channel.

\label{sec:Model}

This reasoning is, however, imprecise as it assumes that one can use $\Pi_{\rm pp} \propto \log^2{T_0/T}$ in a situation when
 the total momentum of the two intermediate fermions in the crossed diagram is non-zero.
  We have explicitly evaluated the
   renormalization of $U$ from the two-loop crossed diagram in Fig.\ref{fig:Stoner}(c) and found
 \begin{equation}
U_{\rm eff} = U \left(1 - b \lambda_{0} \log\frac{T_{0}}{T}\right),
\label{n_2}
\end{equation}
where $b\approx1.88$. The details of this analysis are summarized in Appendix A, where we determine the coefficient $b$ numerically
and show analytically that the naively expected
 $\log^{2}\frac{T_{0}}{T}$
contribution
   vanishes due to a cancellation of different
 terms  that
 individually  scale as $\log^{2}\frac{T_{0}}{T}$.
 
It is {\it a-priori}
unclear how to re-sum the diagrams
 for $U_{\rm eff}$ 
 and whether it is even justified to
  restrict with
  maximally crossed diagrams.  If we assume that higher-order crossed diagrams form a geometrical series, we find
   $U_{\rm eff} = U/(1 + b\lambda_0 \log\left(T_0/T\right))$, which replaces Eq.~\eqref{eq:Stoner_naive}
    for 
    the condition for the Stoner instability by
 \begin{equation}
\frac{\lambda_{0}\log\frac{T_{0}}{T}}{1+\lambda_{0}b\log\frac{T_{0}}{T}}=1 .\label{eq:stoner_better}
\end{equation}
Since $b>1$, one still finds that there
  is
  no Stoner instability. 
   This argument is, however,  a suggestive one, and 
   whether there
is a Stoner instability at a single VH point remains an open question.
In what follows,
we will assume that no ferromagnetic instability takes place
and analyze a potential pairing instability.

\section{Pairing instability}

We now 
show
that there is a pairing
 instability for the cases of both, the ordinary and the extended VH point.
 In order to theoretically detect it  one has to go
  beyond the usual 
   one-loop 
   renormalization  group
  treatment
   and include not only the leading logarithms, which cancel out, as we will see, but also the subleading ones. 
    We will show that the pairing mechanism in our case is 
     of the Kohn-Luttinger type, but is nevertheless
     different from the conventional Kohn-Luttinger 
  scenario. 
  In the latter, the screening of a repulsive Hubbard-type interaction $U$  generates an attractive  pairing interaction in
    the spin-triplet channel with   dimensionless coupling constant
    \begin{equation}
        g\equiv \lambda_0^2 = \frac{m^2 U^2}{4\pi^4},
        \label{eq:coupl_const_g}
    \end{equation}
 which  gives rise to a BCS-type pairing instability.  For an ordinary VH point, 
  this 
  would give rise to $T_c \sim T_0 e^{-1/\sqrt{g}}$  originating from
   $g \log^2 (T_0/T_c) =1$, where one logarithm is a Cooper one and  the other is due to the VH singularity in the density of states; we remind that $T_0$ was introduced below Eq.~\eqref{eq:TcVH}.
  Such a result holds for a Fermi surface without VH points when the attractive pairing interaction
    is a logarithmically singular function of the frequency transfer\cite{Son1999,Chubukov2005}.
In our case, the attraction in the spin-triplet channel does appear due to screening by particle-hole pairs and is of order $U^2$. However, the effective pairing interaction  $\Gamma_{\alpha \beta\gamma \delta} \left(\boldsymbol{k},-\boldsymbol{k},\boldsymbol{p},-\boldsymbol{p}\right)$, where   $\boldsymbol{k}$ and $\boldsymbol{p}$ are relative to the VH point, is strongly momentum dependent in the triplet channel and
is reduced when one momentum is much smaller than the other one. 
 This 
 effectively eliminates the Cooper logarithm leaving only  the one from the density of states. As a result,  we will find that $T_c$ is given by Eq.~\eqref{eq:TcVH}.  The same holds for a higher-order VH point. In this case, there is no exponential dependence of $T_c$ on $g$ but $T_c$ is still  reduced compared to that in the 
  conventional 
  Kohn-Luttinger scenario.

\begin{figure}
    \centering
    \includegraphics[width=\columnwidth]{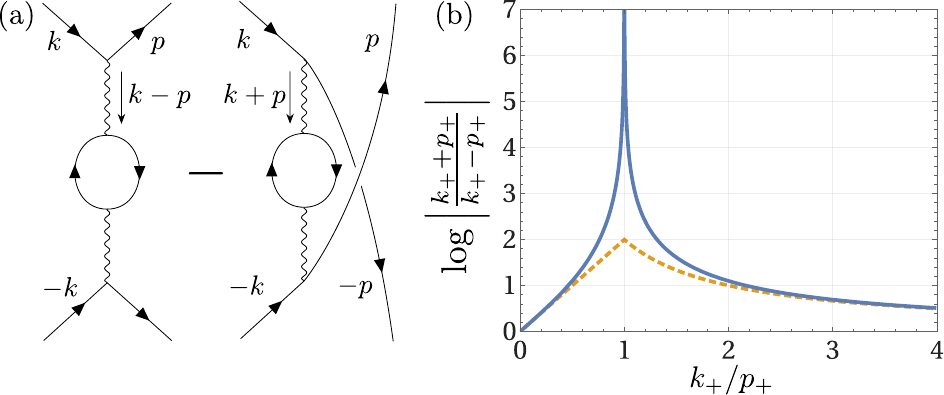}
\caption{(a) Pairing interaction $\Gamma_{\alpha\beta\gamma\delta}\left(\boldsymbol{k},\boldsymbol{p}\right)$ of Eq.~\eqref{eq:vpairing_vertex_ph} 
 at second order in $U$, 
  dressed by 
  particle-hole excitations. 
 Solid lines stand for fermions, which give rise to the bubble $\Pi_{\rm ph}$. The wiggly line stands for the local interaction $U$. (b) The 
  function $\log{\left|\frac{k_+ + p_+}{k_+ -p_+}\right|}$, which determines the triplet component of 
 $\Gamma_{\alpha\beta\gamma\delta} (\boldsymbol{k},\boldsymbol{p})$ for $k_-= p_- =0$, 
 as a function of $k_+/p_+$. The interaction gets weak whenever one of the two momenta is small. The blue solid line is the 
  actual function, while the orange dashed line is an approximate expression 
  based on Eq.~\eqref{eq:Pisplit}.}
    \label{fig:pairing_int}
\end{figure}

A generic recipe for the analysis of potential pairing mediated by nominally repulsive electron-electron interaction  is to consider an irreducible pairing vertex
\begin{equation}
 \Gamma_{\alpha\beta\gamma\delta}\left(\boldsymbol{k},\boldsymbol{p}\right)=\Gamma_{\alpha \beta\gamma \delta} \left(\boldsymbol{k},-\boldsymbol{k},\boldsymbol{p},-\boldsymbol{p}\right)
\end{equation}  instead of the bare $U$ because  $\Gamma$  rather than $U$ appears in the gap equation~\cite{AGD1963}.
The irreducible pairing vertex is the anti-symmetrized interaction with zero 
 total
 incoming and outgoing momenta, dressed by the renormalizations outside of the particle-particle channel, i.e. by  processes which in  a diagrammatic representation have no cross-sections with two fermionic propagators  with opposite momenta.
To second order in $U$, the static irreducible vertex takes the form
~\cite{Maiti2014}
\begin{eqnarray}
\Gamma_{\alpha\beta\gamma\delta}\left(\boldsymbol{k},\boldsymbol{p}\right) & = & U\left(\delta_{\alpha\beta}\delta_{\gamma\delta}-\delta_{\alpha\delta}\delta_{\beta\gamma}\right)\nonumber \\
 & + & U^{2}\Pi_{{\rm ph}}\left(\boldsymbol{k}+\boldsymbol{p}\right)\delta_{\alpha\beta}\delta_{\gamma\delta}\nonumber \\
 &-&U^{2}\Pi_{{\rm ph}}\left(\boldsymbol{k}-\boldsymbol{p}\right)\delta_{\alpha\delta}\delta_{\beta\gamma}.\label{eq:vpairing_vertex_ph}
\end{eqnarray}
The underlying processes are shown in Fig.~\ref{fig:pairing_int}, where
 $\Pi_{{\rm ph}}\left(\boldsymbol{k}\right)$ is the static particle-hole
  polarization bubble.

 The restriction to second order in $U$ may seem questionable as in the
 previous section we argued that the coupling in the particle-hole channel is reduced, to the extent 
 that no Stoner instability takes place.
We will show, however,
 that the typical momenta
 $\boldsymbol{k}$ and $\boldsymbol{p}$,
 responsible for pairing, 
are 
comparable   
to the cutoff $\Lambda$.
For such momenta, the suppression of the coupling by crossed diagrams is small and can be neglected.

  \subsection{Pairing at the ordinary VH point}
The gap equation for the ordinary VH point takes the conventional form
\begin{equation}
\Delta_{\alpha\beta}\left(\boldsymbol{k}\right)=-\int_{\boldsymbol{p}}\frac{\tanh\left(\frac{\varepsilon_{\boldsymbol{p}}}{2(1+Z_{\boldsymbol{p}})T}\right)\Gamma_{\alpha\gamma\beta\delta}\left(\boldsymbol{k},\boldsymbol{p}\right)}{2\varepsilon_{\boldsymbol{p}}\left(1+Z_{\boldsymbol{p}}\right)}\Delta_{\gamma\delta}\left(\boldsymbol{p}\right).
\end{equation}
Here 
$Z_{\boldsymbol{p}}$ is the inverse quasi-particle weight,
related to the fermionic self-energy by $\Sigma\left(\boldsymbol{k},\omega\right)=-i\omega Z_{\boldsymbol{k}}$. 
It is convenient to rotate the
coordinate system
by $\pi/4$ and introduce
 $k_{\pm}=\frac{1}{\sqrt{2}}\left(k_{x}\pm k_{y}\right)$ and
 $p_{\pm}=\frac{1}{\sqrt{2}}\left(p_{x}\pm p_{y}\right)$.
 The Fermi surface around a VH point is specified by either $p_+=0$ or $p_-=0$, 
 see Fig.~\ref{fig:fermisurface}(c).
 In these notations~\cite{Dzyaloshinskii1996,Menashe1999}
\begin{equation}
Z_{\boldsymbol{p}}=2g\log(2)\log\frac{\Lambda}{\left|p_{+}\right|}\log\frac{\Lambda}{\left|p_{-}\right|},
\label{eq:ZID}
\end{equation}

  The gap equation can be split into two decoupled equations for the singlet and triplet components, respectively,  by expressing the gap function as
\begin{equation}
\Delta_{\alpha\beta}\left(\boldsymbol{k}\right)=\Delta_{s}\left(\boldsymbol{k}\right)i\sigma_{\alpha\beta}^{y}+\boldsymbol{\Delta}\left(\boldsymbol{k}\right)\cdot\left(i\sigma^{y}\boldsymbol{\sigma}\right)_{\alpha\beta}.
\end{equation}
The bare interaction $U$ shows up only in the singlet channel, the dressed $\Gamma_{\alpha\beta\gamma\delta}\left(\boldsymbol{k},\boldsymbol{p}\right) $ of Eq.~\eqref{eq:vpairing_vertex_ph}has both, singlet and triplet components.  One easily finds that  there is no solution for $\Delta_s \left(\boldsymbol{k}\right)$ in
the singlet channel, because the dressed pairing vertex  remains repulsive.
In the triplet channel, the gap equation takes the form
\begin{eqnarray}
\Delta_{i}\left(\boldsymbol{k}\right)&=&-U^{2}\int_{\boldsymbol{p}}\frac{\tanh\left(\frac{\varepsilon_{\boldsymbol{p}}}{2(1+Z_{\boldsymbol{p}})T}\right)}{2\varepsilon_{\boldsymbol{p}}(1+Z_{\boldsymbol{p}})}\Delta_{i}\left(\boldsymbol{p}\right) \nonumber \\
& \times & \left(\Pi_{{\rm ph}}\left(\boldsymbol{k}+\boldsymbol{p}\right)-\Pi_{{\rm ph}}\left(\boldsymbol{k}-\boldsymbol{p}\right)\right)
,\label{eq:gap_eq_in_triplet}
\end{eqnarray}
with $i=x,y,z$. In what follows we choose an arbitrary quantization
axis in spin space and drop the index $i$. We get back to this issue
when we discuss superconducting fluctuations.

In 
 terms of  $k_\pm$ and $p_\pm$
 the pairing vertex is the sum of two terms
\bea
&& \Pi_{{\rm ph}}\left(\boldsymbol{k}+\boldsymbol{p}\right)-\Pi_{{\rm ph}}\left(\boldsymbol{k}-\boldsymbol{p}\right) \nonumber \\
&& = \frac{m}{2\pi^2}  \left(\log\frac{|k_{+}- p_{+}|}{|k_{+}+ p_{+}|} + \log\frac{|k_{-}- p_{-}|}{|k_{-}+ p_{-}|}\right).
\label{eq:PiVH}
\eea
 We see that one of the terms  vanishes when either $k_+ =0$ or $k_{-}=0$. This allows to search for
 $\Delta (\boldsymbol{k})$ which depends only on one of the coordinates, i.e.,   $\Delta\left(k_{+}\right)$ or $\Delta\left(k_{-}\right)$. Even if
there 
 exist 
 more complicated solutions, finding a  solution
of this kind is enough to establish a lower bound for  the superconducting
$T_c$.  For definiteness, below we consider
$\Delta (\boldsymbol{k}) =\Delta\left(k_{+}\right)$
 and $\Delta (\boldsymbol{p}) =\Delta\left(p_{+}\right)$.
Under
  this assumption we can perform the integration over 
  $p_-$ 
  at the outset and obtain
  \begin{equation}
    \int_{-\Lambda}^{\Lambda}\frac{dp_{-}}{2\pi}\frac{\tanh\left(\frac{p_{+}p_{-}}{2mT}\right)}{\frac{2p_{+}p_{-}}{m}\left(1+Z_{\boldsymbol{p}}\right)}=\frac{m}{2\pi\left|p_{+}\right|}K\left(p_{+}\right).
    \label{eq:lin_gap_full}
\end{equation}
The function
\begin{eqnarray}
K\left(p_{+}\right)
 & \approx & \frac{\log\left(1+2g\log(2)
 \log\frac{\Lambda}{p_{+}}\log\frac{p_{+}\Lambda}{2mT}\right)}{2g\log(2)\log\frac{\Lambda}{p_{+}}}
\label{ee_4}
\end{eqnarray}
determines the effective density of states 
 for momenta transverse to the Fermi surface.
The expression is valid 
 when $\frac{p_{+}\Lambda}{2mT}\gg 1$,
setting  the temperature dependent lower cutoff 
 for $p_+$ at  $2mT/\Lambda$.
 The upper cutoff is at $p_+ = \Lambda$.
For $g\rightarrow0$,
 $K\left(p_{+}\right)\rightarrow\log\frac{p_{+}\Lambda}{2mT}$.

\begin{figure}
    \centering
 \includegraphics[width=.9\columnwidth]{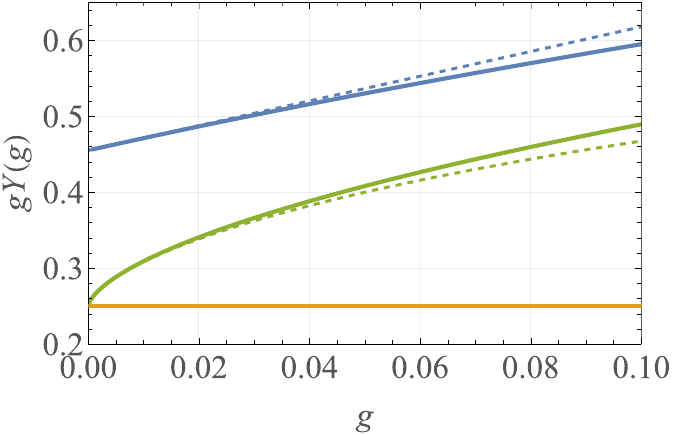}
    \caption{Variation of $gY\left(g\right)=g\log\frac{T_{0}}{T_{c}(g)}$ 
     with the dimensionless coupling constant $g$. Blue curves 
      are 
      the full solution of the linearized gap equation \eqref{eq:gapVH}
      with self-energy included.
      The blue dashed line is a fit to $(1+\mu g)/\gamma$ with 
      $\gamma$ and $\mu$ given in Eq.~\eqref{eq:gmmu}.  This dependence of
     $gY(g)$ leads to  Eq.~\eqref{eq:TcVH} for  $T_c(g)$.  The green 
      lines
     are
      the solutions of the gap equation without self-energy,
       which we discuss in section~\ref{analytic_noself_energy}.
        The solid green line is the full numeric solution while the dashed line is the analytic solution of 
        an approximate Eq.~\eqref{nn_8}. The orange line is the 
         zero-order approximate result, 
         Eq.~\eqref{nn_7}. Recall that  larger $Y(g)$ correspond to smaller $T_c$.    }
    \label{fig:Yofg_VH}
\end{figure}

\begin{figure}
    \centering
 \includegraphics[width=\columnwidth]{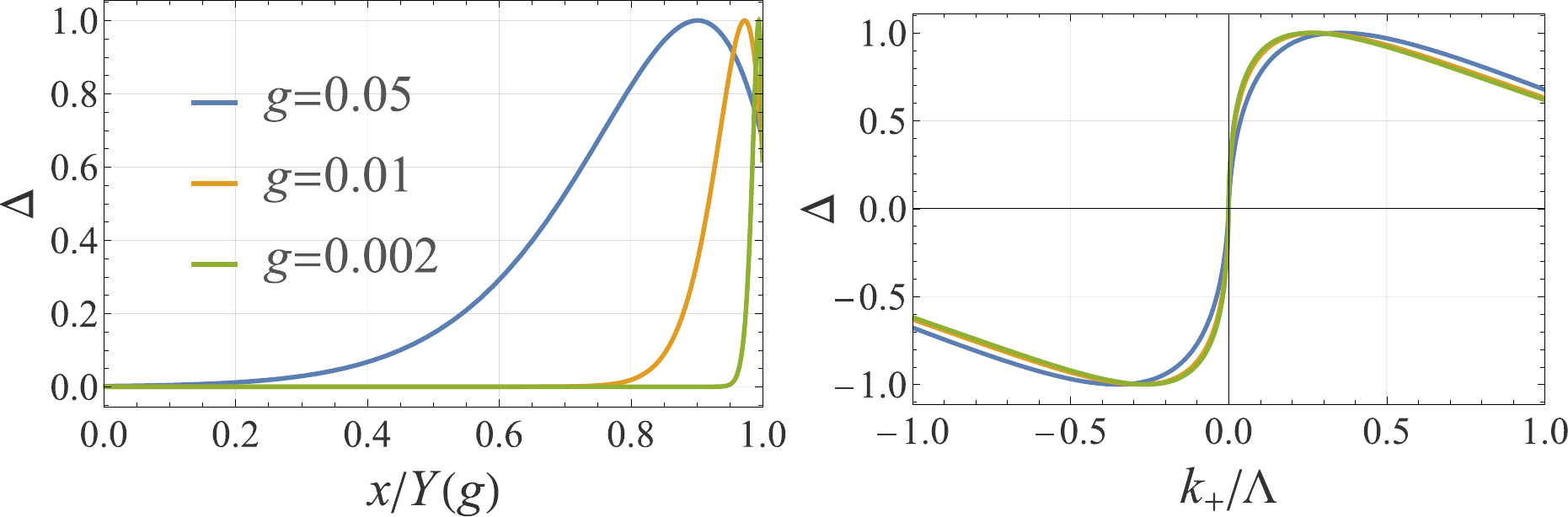}
    \caption{Momentum dependence of the superconducting gap function, obtained from the solution of the linearized gap equation \eqref{eq:gapVH}
    with effective density of states $ K\left(\bar{p}\right)$ 
     given by Eq.~\eqref{eq:Kdimless}. 
     Such a gap function develops infinitesimally below $T_c$. 
     The gap function is 
      normalized to its maximal value. Left panel:
    $\Delta(x)$ 
     as a function of $x/Y(g)$, where $x=\log\left(\frac{\Lambda k_{+}}{2mT}\right) $ 
      and $Y(g) = \log{T_0/T_c (g)}$. Right panel: $\Delta(k_+)$ as function of $k_+/\Lambda$. Notice the non-monotonic behavior of the gap function with a steep slope at small momenta. }
    \label{fig:gap_VH}
\end{figure}

 It is convenient to 
 introduce dimensionless variables $\bar{p}=\frac{\Lambda p_{+}}{2mT}$ and similar for
$\bar{k}$,
 and $\bar{\Lambda}=\frac{\Lambda^{2}}{2mT}$.
   In these variables  the gap equation takes the form
\begin{equation}
    \Delta\left(\bar{k}\right)=-g\int_{1}^{\bar{\Lambda}}\frac{d\bar{p}}{\bar{p}}K\left(\bar{p}\right)\log\left|\frac{\bar{k}-\bar{p}}{\bar{k}+\bar{p}}\right|\Delta\left(\bar{p}\right),
    \label{eq:gapVH}
\end{equation}
with
\begin{equation}
    K\left(\bar{p}\right)=\frac{\log\left(1+2g\log(2)\log\left(\bar{\Lambda}/\bar{p}\right)\log\bar{p}\right)}{2g\log(2)\log\left(\bar{\Lambda}/\bar{p}\right)}.
    \label{eq:Kdimless}
\end{equation}
We solved this equation numerically 
 and obtained $T_c (g)$ and $\Delta ({\bar k})$.
 In Fig.~\ref{fig:Yofg_VH} we plot $g Y\left(g\right)$
 with
 \begin{equation}
  Y\left(g\right)\equiv  \log\frac{T_{0}}{T_{c}}  
 \end{equation}
  as function of the coupling constant
$g$.
We find that at small $g$ the behavior is well described by a linear relation
$g \log\frac{T_{0}}{T_{c}}  = (1+\mu g)/\gamma$
with
\begin{eqnarray}
    \gamma &\approx& 2.197 \nonumber \\
    \mu &\approx &  3.515.
    \label{eq:gmmu}
\end{eqnarray}
 This yields  the transition temperature given in Eq.~\eqref{eq:TcVH}.
 In Fig.~\ref{fig:gap_VH} we show the momentum dependence of the gap function $\Delta(\bar{k})$ 
  extracted from the numerical solution. $\Delta(\bar{k})$ is odd under $\bar{k}\rightarrow -\bar{k}$ as required for an odd-parity triplet state. The momentum dependence is non-monotonic with a maximum 
  at an intermediate momentum, 
which scales with 
$\Lambda$ but numerically is much smaller than $\Lambda$. 
   The linear dependence of $\Delta(\bar{k})$ on
    $\bar{k}$ holds 
at even smaller momenta below the maximum. 

To gain physical insight and better interpret our numerical findings, we next perform an analytic analysis  of the gap equation.
We   obtain two   approximate solutions,  which qualitatively reproduce the full numerical one and provide transparent  insights into  
 the  key aspects of the pairing instability.

\subsubsection{ An approximate  analytical  solution assuming a constant density of states}

We begin by analyzing the gap equation under the simplifying assumption that 
 relevant ${\bar p}$ are of order ${\bar \Lambda}$ ($p_+ \sim \Lambda$). In this situation, one can expand Eq.~\eqref{ee_4} in  $g$  and obtain a momentum-independent density of states $K(p_+) \approx \log{{\bar \Lambda}}$, which can be pulled out of the integral over ${\bar p}$ in Eq.~\eqref{eq:gapVH}. 
 The gap equation then  reduces  to
 \begin{equation}
\Delta\left(\bar{k}\right)= g^* \int_{1}^{\bar{\Lambda}}
\frac{d\bar{p}}{\bar{p}} \log\left|\frac{\bar{k} +\bar{p}}{\bar{k} - \bar{p}}\right|\Delta\left(\bar{p}\right),
\label{ch_4}
\end{equation}
with effective coupling constant
\beq
 g^* = g \log{\bar{\Lambda}}.
\label{ch_5}
\eeq
 As a further simplification, we 
  follow Refs. \cite{Chubukov2005,Chubukov_gamma}
  and 
 split the integral over ${\bar p}$  
  into the regimes $\bar{p}<\bar{k}$ and
$\bar{p}>\bar{k}$ and in each regime approximate
\begin{equation}
\log\left|\frac{\bar{k}-\bar{p}}{\bar{k}+\bar{p}}\right|
 \approx 
 \left\{ \begin{array}{cc}
-2\frac{\bar{p}}{\bar{k}} & {\rm if}\,\,\bar{p}\ll\bar{k}\\
-2\frac{\bar{k}}{\bar{p}} & {\rm if}\,\,\bar{k}\ll\bar{p}
\end{array}\right. .
\label{eq:Pisplit}
\end{equation}
 We then obtain
\begin{eqnarray}
\Delta \left(\bar{k}\right)= 2g^*  \left[ \frac{1}{{\bar k}} \int_{1}^{{\bar k}} d {\bar p} \Delta \left(\bar{p}\right)
+ {\bar k}  \int_{{\bar k}}^{\bar{\Lambda}}   \frac{d {\bar p}}{\bar{p}^2} \Delta\left(\bar{p}\right)\right].
\label{ch_7}
\end{eqnarray}
Differentiating w.r.t. ${\bar k}$ we reduce Eq.~\eqref{ch_7} to  
 a second order differential equation
\begin{equation}
\bar {k}^2 \Delta'' \left(\bar{k}\right) + {\bar k} \Delta' \left(\bar{k}\right)  + (4 g^* -1) \Delta \left(\bar{k}\right) =0,
\label{ch_8}
\end{equation}
 with UV and IR  boundary conditions
\bea
\bar{\Lambda}\Delta' \left(\bar{\Lambda}\right) &=& - \Delta \left(\bar{\Lambda}\right),
%
\label{ch_9} \\
\Delta' \left(1\right) &=&  \Delta \left(1\right),
\label{ch_10}
\eea
imposed by the original integral equation. For $4 g^* <1$ (i.e., at higher temperatures), the  solution of Eq.~\eqref{ch_8} is
\beq
\Delta \left(\bar{k}\right) =\Delta_0 \left( {\bar k}^{\sqrt{1-4g^*}} + 
e 
{\bar k}^{-\sqrt{1-4g^*}}\right),
\label{ch_11}
\eeq
where $e$ is a free parameter.  
We verified that 
one cannot choose $e$ to 
 satisfy both  boundary 
 conditions. Hence,  there is no solution for $g^* <1/4$.
For  $4g^{*}>1$ the  solution is
\begin{equation}
\Delta\left(k\right)=\Delta_{0}\cos\left(\sqrt{4g^{*}-1}\log k+\phi\right),
\end{equation}
where $\phi$ is a free parameter. The 
IR boundary condition specifies  $\phi$ to be the solution of 
 $\tan\phi=-\left(4g^{*}-1\right)^{-1/2}$,
 while the UV
 boundary 
 condition yields
\begin{equation}
  \sqrt{4g^{*}-1}\log\bar{\Lambda}=l\pi-2\arctan\left(\sqrt{4g^{*}-1}\right),
\end{equation}
with integer $l$. 
 Using ${\bar \Lambda} = \Lambda^2/(2m T)$, we find that this equation
 determines a
  discrete set of critical temperatures $T_c (l)$. 
The largest $T_c$  corresponds to $l=1$  and is the solution of
 $\sqrt{4g^{*}-1}\approx\pi /\log\bar{\Lambda}$.
Solving at small $g$, we obtain 
\beq
T_c \approx T_0 e^{-1/(4g)}.
\label{nn_7}
\eeq 
This is similar to the  expression given in Eq.~(\ref{eq:TcVH}), but with $\gamma =4$ and $\mu =0$. 
 This solution is the orange line in Fig.~\ref{fig:Yofg_VH}. 
 While 
 this approximation
 does not reproduce our numerical findings quantitatively, it does 
  capture the two key features of the pairing instability of our problem.
    First, the Cooper logarithm is suppressed
     because the pairing interaction in Eq.~\eqref{eq:PiVH}
     gets suppressed when an internal momentum is larger than an external one and vice versa. 
     Taken alone, 
     this suppression 
      would 
       impose 
       a threshold value for pairing, i.e. $T_c$ 
       would be nonzero only 
       for $g$ larger than some critical value. 
       Second,
       the large density of states, encoded in the phase space of transverse momenta that determine the function $K({\bar k})$  
         in 
         Eq.~\eqref{eq:Kdimless}, compensates for the weak pairing: 
          it is the effective coupling constant $g^*
           = g \log\left[\Lambda^{2}/(2mT)\right]$ of Eq.~\eqref{ch_5} that must reach a threshold.
           Because 
           $g^*(T\rightarrow 0)\rightarrow \infty$, the threshold condition is satisfied at a finite $T_c$ for any $g$. The combination of the two effects yields $T_c$, which has  the same form as the BCS expression, albeit for a different reason.

There 
  is a certain analogy between 
  the solution for $T_c$ and the gap function in 
  our model and  in the model with the  singular dynamical interaction between fermions, $\chi (\Omega_m) \propto 1/|\Omega_m|^\gamma$ (the $\gamma$ model)~\cite{Chubukov_gamma,Schmalian_syk,Wang_syk,Classen_syk,Chubukov_gamma_a}. In both cases, the integral equation for the gap function can be reduced to the differential equation with a marginal kernel, whose solution  yields a power-law dependence of the gap function $x^{\pm a}$ ($x$ is a momentum in our case and a frequency in the $\gamma$ model), where $a$ depends on $T$ in our case
($a = \sqrt{1-4g^*}$) 
and on $\gamma$ in the $\gamma-$model.
 As long as the exponent $a$ is real 
  the potential solution $\Delta (x) =x^a  + b x^{-a}$ does not satisfy the two boundary conditions, hence there is no superconductivity.
 A non-zero solution  develops when
the exponents $\pm a$ merge.  In this case, besides a constant, there appears the second solution $\log x$. The candidate solution $\Delta (x) =1 + b \log{x}$ satisfies the two boundary conditions, which
  fix the value of $b$, and  hence is the actual solution of the linearized gap equation.
  This implies that $a=0$  is the condition for $T_c$ Also, in both cases
  the solution of the linearized gap equation exists even at $4g^* >1$,
    as the end point of the infinite set of solutions of the non-linear gap equation.

\subsubsection{Analytic solution for momentum-dependent density of states}
\label{analytic_noself_energy}
  We now advance to the next level and drop the assumption that the density of states $K(p_+)$  can be approximated as momentum-independent.  
  We still expand the density of states to first order in $g$ (and, by doing this, neglect the self-energy), but keep $K ({\bar p}) = \log{{\bar p}}$ instead of  $\log{\bar \Lambda}$ (we remind that 
  $\bar{p}=\frac{\Lambda p_{+}}{2mT}$).
The gap equation takes the form
\begin{equation}
\Delta\left(\bar{k}\right)=- g \int_{1}^{\bar{\Lambda}}\frac{d\bar{p}}{\bar{p}}\log\bar{p}\log\left|\frac{\bar{k}-\bar{p}}{\bar{k}+\bar{p}}\right|
\Delta\left(\bar{p}\right).\label{eq:integral_eq}
\end{equation}
The  lower momentum cutoff remains  at ${\bar p} =1$.
 We again 
 split the integral over ${\bar p}$ into the regions $\bar{p} <\bar{k}$ and $\bar{p} >\bar{k}$  and use Eq.~\eqref{eq:Pisplit}.
 We 
 then
  obtain 
\begin{equation}
\Delta\left(\bar{k}\right)=2
g
\int_{1}^{\bar{k}}\frac{d\bar{p}}{\bar{k}}\log\bar{p}\Delta\left(\bar{p}\right)+2 g \bar{k}\int_{\bar{k}}^{\bar{\Lambda}}\frac{d\bar{p}}{\bar{p}^2}\log\bar{p}\Delta\left(\bar{p}\right).
\end{equation}
In logarithmic variables $x=\log\bar{p}$ and $y=\log\bar{k}$  this
becomes
\begin{equation}
\Delta\left(y\right)=2g
\int_{0}^{y}dxxe^{x-y}\Delta\left(x\right)+2 g
\int_{y}^{Y}dxxe^{y-x}\Delta\left(x\right), 
\label{ch_19}
\end{equation}
 where $Y=\log\bar{\Lambda}$.
Differentiating twice over $y$, we find
 that
 (\ref{ch_19})  is equivalent to the second
order differential equation
\begin{equation}
\frac{d}{dy}e^{-2y}\frac{d}{dy}e^{y}\Delta\left(y\right)=-4 g
e^{-y}\Delta\left(y\right),
\label{nn_6}
\end{equation}
with UV and IR
 boundary conditions:
\begin{eqnarray}
\left.\frac{d\Delta}{dy}\right|_{Y} & = & -\left.\Delta\right|_{Y},\nonumber \\
\left.\frac{d\Delta}{dy}\right|_{0} & = & \left.\Delta\right|_{0}.
\end{eqnarray}
The 
differential equation 
(\ref{nn_6}) 
 equals to
\begin{equation}
\Delta''=\left(1-4 g y\right)\Delta,
\end{equation}
 This last equation 
 is equivalent to the Schrödinger equation in a linear potential
$V\left(y\right)=1-4\lambda^{2}y$ 
 for 
 zero eigenvalue, $E=0$.
The 
"coordinate" varies between 
 $y=0$ 
and
 $y=Y=\log\frac{\Lambda^{2}}{2mT}$.
If $4\lambda^{2}Y<1$ the potential never becomes
negative and getting a 
solution with zero eigenvalue is not possible. However, for
$4g Y>1$, i.e. $T<\frac{\Lambda^{2}}{2m}e^{-\frac{1}{4g}}$
a non-zero solution becomes 
a possibility.
Introducing 
  $z=\left(1-4g y\right)/\left(4g\right)^{2/3}$, we re-express 
the differential equation 
 as 
  $\Delta''=z\Delta$, whose solution
 is a linear combination of the  Airy functions ${\rm Ai}\left(z\right)$ and ${\rm Bi}\left(z\right)$.
In terms of the original variable $y$,
\begin{equation}
\Delta\left(y\right)= \Delta_0 \left(
{\rm Ai}\left(\frac{1-4g
y}{\left(4 g
 \right)^{2/3}}\right)+
 c
{\rm Bi}\left(\frac{1-4 g
y}{\left(4g
\right)^{2/3}}\right) \right).
\label{eq:gap_ordVH}
\end{equation}
At $y=0$ and small $g\ll$1 the argument of the Airy functions
is large and positive. Using the  asymptotic expressions of ${\rm Ai}$ and ${\rm Bi}$,
 we find
  from the boundary condition at $y=0$ that
 $c \approx (g/4) e^{-\frac{1}{3g}}$
  is exponentially small.
 Since both functions are comparable in
magnitude at the UV boundary condition at $y = Y$,
we can safely neglect ${\rm Bi}$ in (\ref{eq:gap_ordVH}),
 i.e., approximate the gap function by
 \beq
 \Delta\left(y\right)= \Delta_0
{\rm Ai}\left(\frac{1-4g  y}{\left(4g\right)^{2/3}}\right).
\label{ch_20}
\eeq

The UV boundary condition at $y=Y$ 
 then
 yields
\begin{equation}
{\rm Ai}\left(\frac{1-4g
Y}{\left(4g
\right)^{2/3}}\right)=\left(4g\right)^{1/3}{\rm Ai'}\left(\frac{1-4g
Y}{\left(4g
\right)^{2/3}}\right).
\label{ch_21}
\end{equation}
This condition determines the critical temperature through the $T$-dependence
of $Y(T_c) = \log{\left(\Lambda^2/(m T_c)\right)}$.
 For small $g$, the 
 solutions of ${\rm Ai} (z) = (4 g)^{1/3} {\rm Ai}' (z)$ 
  are
 $z \approx z_n$, where $z_n$ is the n-th zero of
 the Airy function. The largest $T_c$ corresponds to the first zero at
$z_{0}\approx-2.338$.   For a more accurate analysis we expand Eq.~\eqref{ch_21} around $z = z_0$ as
${\rm Ai}\left(z\right)\approx{\rm Ai}'\left(z_{0}\right)\left(z-z_{0}\right)$
and ${\rm Ai}'\left(z\right)\approx{\rm Ai'}\left(z_{0}\right)+{\cal O}\left(\left(z-z_{0}\right)^{2}\right)$.
We then obtain 
$1-4 g Y-z_{0}\left(4 g \right)^{2/3}=4 g$ that yields 
 \beq
 T_c = T_0  \exp\left({-\frac{1 + |z_{0}|\left(4 g \right)^{2/3}-4g}{4g}}\right). 
\label{nn_8}
\eeq
 We plot this $T_c$ in Fig.~\ref{fig:Yofg_VH}, where we also compare it with the full numerical solution of the gap equation \eqref{eq:integral_eq} without self energy corrections (dashed and solid green lines, respectively). 
 We verified that the numerical solution without the self-energy does reproduce the $g^{2/3}$ dependence in the exponent for $T_c$, as in Eq.~\eqref{nn_8}. 
Further  comparing this equation 
with Eq.~\eqref{nn_7} we see that the leading exponential dependence in both formulas is the same  $e^{-1/(4g)}$, i.e. $\gamma =4$, larger in the numerics,  but 
 the additional $O(g^{2/3})$ term in the numerator in  Eq.~\eqref{nn_8} 
 effectively reduces $\gamma$. 
    We show in the next subsection that 
    the self-energy corrections
     change 
     the functional form of this term from $O(g^{2/3})$ to $O(1)$ such that the value of $\gamma$ indeed gets reduced, 
     in agreement with the  numerical solution of the
      full 
      gap equation.

The analytical $\Delta (k_+)$ from Eq.~\eqref{ch_20} can be re-expressed as 
 \begin{equation}
\Delta\left(k_{+}\right)= \Delta_{0}
{\rm Ai}\left(z_0 + (4g)^{1/3} \left(1 - \log\frac{k_{+}}{\Lambda}\right)\right).
\label{ch_24}
\end{equation}
\begin{figure}
    \centering
    \includegraphics[width=\columnwidth]{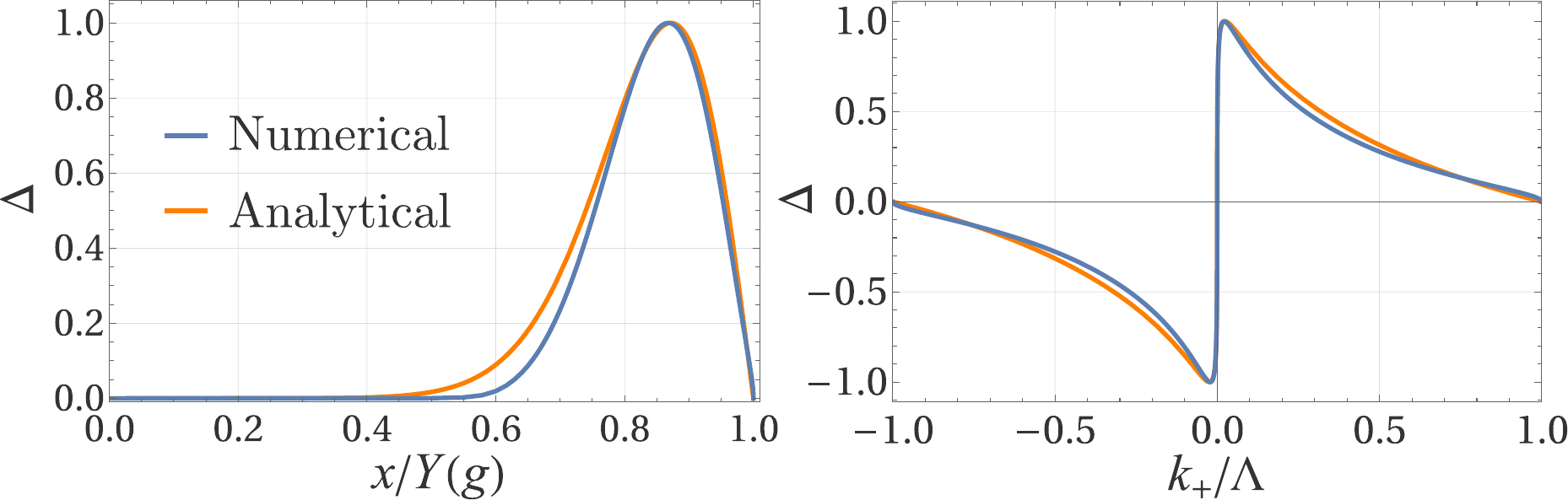}
    \caption{Momentum dependence of the triplet gap function for the ordinary VH point,
     obtained by solving the gap equation without self-energy corrections
     in the density of states kernel $K ({\bar p})$. 
      The gap function is normalized to its maximal value.
      The two plots are 
    on a logarithmic scale with $x=\log\left(\frac{\Lambda k_{+}}{2mT}\right)$ (left panel)  
    and 
    on a linear scale (right panel). The dimensionless coupling constant is $g=0.01$.  The blue line is obtained by solving Eq.~\eqref{eq:integral_eq} numerically, and the orange line is the solution of Eq.~\eqref{eq:gap_ordVH}. 
    Notice that the region where the gap function is linear in momentum is narrower than in Fig.~\ref{fig:gap_VH}, which includes self-energy corrections.
    }
    \label{fig:triplet_gap}
\end{figure}
We remind that this formula is valid for  $k_{+}>2mT_{c}/\Lambda$, where our computational procedure holds. 
At smaller $k_+$, we expect $\Delta (k_+)$ to scale linearly with $k_+$. 
 The gap function in Eq.~\eqref{ch_24} has a sharp maximum at 
\begin{equation}
k_{{\rm max}}=\Lambda\exp\left(-\frac{\left|z_{0}\right|-\left|\tilde{z}_{0}\right|-\left(4g\right)^{1/3}}{\left(4g\right)^{1/3}}\right),
\end{equation}
where $\tilde{z}_{0}=-1.019$ (${\rm Ai}'\left(\tilde{z}_{0}\right)=0$).
 In Fig.~\ref{fig:triplet_gap} we compare $\Delta\left(k_{+}\right)$ from (\ref{ch_24}) with the numerical solution of the original, integral gap equation   without self-energy corrections.  The agreement is quite good. 

\subsubsection{
Analytic solution with self-energy included}

In the two previous subsections we expanded the density of states $K(p_+)$ of Eq.~\eqref{ee_4} to first order in  $g$ and, in doing so, neglected the fermionic self-energy.  Such an expansion, however, holds only for $p_+$ very near $\Lambda$ and becomes problematic for somewhat smaller $p_+$, even if comparable to $\Lambda$.  Indeed, in logarithmic variables the density of states is expressed  as
\beq
K(y) = \frac{\log{(1 + 2 \log(2) g y (Y-y))}}{\alpha g (Y-y)}.
\label{ee_8}
\eeq
 Because typical $y$ are comparable to $Y$ and $g Y = O(1)$, the expansion holds only when $Y -y <1$.  In the previous section, we expanded (\ref{ee_8}) to order $g$, but assumed that the solution of the differential equation holds even when $Y-y \geq 1$.  Now we go beyond the linear order in $g$ and demonstrate  that this affects the results and improves the agreement with the numerical finding.   We will not attempt to solve the gap equation with the full $K(y)$ but rather expand it to second order in $g$ and analyze how it affects $T_c$.

Expanding to order $g^2$ and keeping in mind that the relevant $y$ are still close to $Y$, we obtain
\beq
K(y) \approx y - g Y^2 (Y-y) \log{2}.
\eeq 
Following  the same computational steps as in the previous subsection, we obtain the 
differential gap equation
\beq
\Delta''(y)= V_{\rm eff}(y) \Delta (y),
\eeq
with effective potential 
\beq
V_{\rm eff} (y)  = 1-4g K (y) = 1 - 4 g_y (y - y_r)
\eeq
where
\beq
g_y = g + (g Y)^2 \log{2}
\label{ee_9}
\eeq
 and 
 \beq
 y_r = Y \frac{(gY)^2\log{2}}{g + (gY)^2\log{2}}
 \label{ee_10}
\eeq
Introducing $z = y-y_r$, we re-express the differential gap equation as 
\beq
\Delta''(z) = V_{\rm eff}(z) \Delta (z),
\eeq
where
\beq
V_{\rm eff} (z) = 1 - 4 g_y (z).
\eeq
This equation is valid for $y <Y$, i.e., for $z < Y_r$, where
\beq
Y_r = Y - y_r = \frac{Y}{1 + g Y^2\log{2}}.
 \label{ee_11}
\eeq
This is the same equation as in the previous subsection, but with $g_r$ instead of $g$ and $Y_r$ instead of $Y$.
Accordingly, the expression for $T_c$ is 
\beq
Y_r = \frac{1 + |z_0| (4 g_r)^{2/3} - 4 g_r}{4 g_r}.
 \label{ee_12}
\eeq
Using $Y_r = Y g/g_r$, we re-express this relation as
\beq
Y = \frac{1 + |z_0| (4 g_r)^{2/3} - 4 g_r}{4 g}.
 \label{ee_14}
\eeq
Our previous result, Eq.~(\ref{nn_8}), for $T_c$ is reproduced if we set $g_r = g$, i.e. neglect the term with the prefactor $\log{2}$ in (\ref{ee_9}).    We see, however, that the $(gY)^2\log{2}$ term is {\it parametrically} larger than $g$, and hence $g_r$ is  larger than $g$. Substituting the expression for $g_r$ into 
(\ref{ee_14}) and keeping only the leading term in $g_r$, we obtain the equation for $a = g Y$ in the form
\beq
4a = 1 + |z_0| (4 \log{2})^{2/3}  a^{2/3} - 4 a \log{2}.
\eeq
The solution of this equation is $a \approx 1/1.7544$, i.e., $T_c \sim T_0 e^{-1/\gamma_r g}$ with $\gamma_r = 1.7544$. Expanding in Eq.~\eqref{ee_12} further to order $g$ we obtain
 \beq
 T_c = T_0 e^{\frac{1+ \mu_r g}{\gamma_r g}},
 \label{ee_15}
 \eeq
 with $\mu_r =  2.092$.  The functional form of (\ref{ee_15}) is the same as extracted from the numerical solution of the gap equation, 
  and the values of $\gamma_r$ and $\mu_r$ are reasonably close to numerical values
  in 
  Eq.~\eqref{eq:gmmu}.  To achieve 
   a 
   better agreement one would have to expand further in $g$.  Still, the expansion of $K (y)$ to order $g^2$ clearly shows that the effect of keeping the self-energy  is effectively a replacement of the $g^{2/3}$ term in the numerator in the expression for $T_c$ in (\ref{nn_8}) by a term of order one. 
  
 \subsubsection{Comparison with Son's model}   
 \label{Sonsmodel}
 It is also instructive to compare our result for $T_c$ with the one for fermions away from VH singularity, but with an attractive  logarithmic interaction.  Such a model was originally solved by Son in the frequency domain\cite{Son1999}, see also Ref.~\cite{Chubukov2005}.
 In momentum space,  the corresponding gap equation  in our notations ${\bar k}= k/\Lambda$  has  the form
\begin{equation}
\Delta\left({\bar k}\right)= \frac{g}{2}  \int_{1}^{\bar{\Lambda}}\frac{d\bar{p}}{\bar{p}}
\log{\frac{\bar{\Lambda}^2}{\left|\bar{k}^2 -\bar{p}^2\right|} \Delta\left(\bar{p}\right)}.
\label{ch_4_1}
\end{equation}
where $\bar{\Lambda} = \Lambda^2/(2 m T)$, as before. We also keep the notation  $g$ for the dimensionless coupling constant.  At a first glance,  the effect of logarithmic interaction is the same as
      from logarithmic density of states at the VH point 
      (Eqn.~\eqref{eq:gapVH}).
      We show, however, that these two problems are rather distinct, because in our case 
      the effective interaction $\log\left|\frac{\bar{k} + \bar{p}}{\bar{k} -\bar{p}}\right|$, while generally of order $O(1)$,  is strongly  reduced when one momenta is much smaller than the other one. This effectively eliminates the Cooper logarithm that emerges through  $d\bar{p}/\bar{p}$ in both
      \eqref{eq:gapVH} and 
      \eqref{ch_4_1}.
 To demonstrate the distinct behavior of these two problems, we split the integral over ${\bar p}$ 
  in
  the r.h.s of (\ref{ch_4_1}) into  regions $\bar{p} <\bar{k}$ and  $\bar{p}>\bar{k}$,  as we did before, 
  introduce logarithmic variables $x = \log{\bar k}$, $y = \log{\bar p}$ as well as $Y=\log\bar{\Lambda}$, and 
  convert (\ref{ch_4_1})  into the 
  the differential equation
\begin{equation}
\Delta''\left(x\right)=-g\Delta\left(x\right)
\label{ch_15}
\end{equation}
with boundary conditions
\begin{eqnarray}
\Delta'\left(0\right) & = & 0, \nonumber \\
\Delta\left(Y\right) & = & 0.
\end{eqnarray}
The solution in terms of the original variable ${\bar k}$ is
\begin{equation}
\Delta ({\bar k}) =  \cos{\left(\sqrt{g} \log{{\bar k}} + \phi\right)}.
\label{ch_17}
\end{equation}
We see that $\Delta ({\bar k})$  oscillates as a function of $\log{{\bar k}}$ for any value of $g$, even infinitesimally small ones.  This is in contrast to our earlier discussion, where an oscillating solution 
holds only for $\lambda$ above the threshold.
The IR boundary condition yields $\phi=l\pi$, with integer $l$, while the UV condition 
 yields
 $\cos\left(\sqrt{g}Y\right)=0$, i.e. $\sqrt{g}Y=\left(2l+1\right)\frac{\pi}{2}$.
The solution 
with $l=0$ 
gives 
 the highest 
 transition temperature 
\begin{equation}
T_{c}=\frac{\Lambda_{0}^{2}}{2m}e^{-\frac{\pi}{2\sqrt{g}}}.
\label{ch_18}
\end{equation}
 Other solutions with $l >0$ yield smaller $T_{c}^{\left(l \right)}=\frac{\Lambda_{0}^{2}}{2m}e^{-\frac{\left(2l+1\right)\pi}{2\sqrt{g}}}$,
Comparing (\ref{ch_18}) to Eq.~(\ref{eq:TcVH}),
we find that  $T_c$  for the single VH point is  smaller as it contains $1/g$ in the exponent as opposed to $1/\sqrt{g}$ in Eq.~(\ref{ch_18}).  The distinction is due to the form of the pairing interaction in the triplet channel. In the case of  pairing at the VH point, the pairing strength alone is too weak to give rise to a Cooper instability. However, the enhanced phase space for scattering, which is a consequence of the logarithmic density of states, compensates for the weak interaction and yields, in the end, a BCS-type expression of the transition temperature.

\subsection{Pairing at the extended VH point}
\label{sec:extended}

Next we analyze extended VH points with  a dispersion relation
given in Eq.~\eqref{eq:extVH} for small but finite power-law exponent
$\epsilon$ that determines the density of states.  
We show in Appendix B that 
the inverse quasi-particle weight 
in this case 
is given by
\begin{equation}
Z_{\boldsymbol{k}}=\frac{2\log2}{4\epsilon^{2}}g
\left(\left(\frac{\Lambda}{k_{+}}\right)^{2\epsilon}-1\right)\left(\left(\frac{\Lambda}{k_{-}}\right)^{2
\epsilon}-1\right),
\label{eq:Z_eps}
\end{equation}
 where the dimensionless coupling constant  
 is now 
\begin{equation}
g=\frac{U^{2}\Lambda^{-4\epsilon}}{\left(4\pi^{2}A\right)^{2}}.
\label{ee_16}
\end{equation}
 Note that 
  this $g$ 
  explicitly depends on the cutoff $\Lambda$. This will
play a role when we   consider the limit where $g\ll \epsilon$.

The gap equation can be written as
\begin{eqnarray}
\Delta\left(k_{+}\right)&=&-\frac{U^{2}\Lambda^{-4\epsilon}}{4\pi Ap_{+}^{1-2\epsilon}}\int_{p_{0}}^{\Lambda}\frac{dp_{+}}{\pi}K\left(p_{+}\right)\Delta_{i}\left(p_{+}\right) \nonumber \\
&\times& \left(\Pi\left(k_{+}+p_{+}\right)-\Pi\left(k_{+}-p_{+}\right)\right)
\label{yy_99}
\end{eqnarray}
where we introduced
\begin{eqnarray}
K\left(p_{+}\right)&=&\frac{4\pi Ap_{+}^{1-2\epsilon}}{\Lambda^{-4\epsilon}}\int_{-\Lambda}^{\Lambda}\frac{dp_{-}}{2\pi}\frac{\tanh\left(\frac{\varepsilon_{\boldsymbol{p}}}{2\left(1+Z_{\boldsymbol{p}}\right)T}\right)}{2\varepsilon_{\boldsymbol{p}}\left(1+Z_{\boldsymbol{p}}\right)}.
\label{yy_9999}
\end{eqnarray}
Here, 
$p_{0}=\frac{T}{A\Lambda^{1+2\epsilon}}$
is the temperature-dependent lower cutoff of the theory. 
 
 For the particle-particle
bubble at small $\epsilon$ we 
 find
\begin{equation}
\Pi\left(k_{+}\right)=\frac{\left|k_{+}\right|^{-2\epsilon}}{8\pi^{2}A\epsilon}.
\end{equation}
This allows 
 us to approximate the pairing kernel as
\begin{equation}
\Pi\left(k_{+}+p_{+}\right)-\Pi\left(k_{+}-p_{+}\right)=\left\{ \begin{array}{cc}
2\Pi'\left(k_{+}\right)p_{+} & {\rm if}\,\,p_{+}\ll k_{+}\\
2\Pi'\left(p_{+}\right)k_{+} & {\rm if}\,\,k_{+}\ll p_{+}
\end{array}\right. \label{eq:kernal}
\end{equation}
with $\Pi'\left(k_{+}\right)=-\frac{1}{4\pi^{2}A}\left|k_{+}\right|^{-(1+2\epsilon)}.$

 Re-expressing the coupling 
 in (\ref{yy_99}) 
 in terms of the dimensionless 
 $g$ from (\ref{ee_16}), 
 splitting the integration over $p_+$ 
 in 
 into the ranges $p_+ < k_+$ and $p_+ > k_+$ and
  using 
   Eq.~(\ref{eq:kernal}), we 
re-express the gap equation as 
\begin{eqnarray}
\Delta\left(k_{+}\right)&=&2g\int_{k_{0}}^{k_{+}}dp_{+}\frac{p_{+}^{2\epsilon}}{k_{+}^{1+2\epsilon}}K\left(p_{+}\right)\Delta_{i}\left(p_{+}\right) \nonumber \\
&+&2g\int_{k_{+}}^{\Lambda}dp_{+}\frac{k_{+}}{p_{+}^{2}}K\left(p_{+}\right)\Delta_{i}\left(p_{+}\right).
\end{eqnarray}
Rescaling the gap function as
\[
\Delta\left(k_{+}\right)={\bar \Delta}\left(\log\frac{k_{+}}{p_{0}}\right)\left(\frac{k_{+}}{\Lambda}\right)^{-\epsilon}
\]
  and introducing again the logarithmic variable
 \beq
 x = \log\frac{k_{+}}{p_{0}},
 \eeq
 we obtain  a Schr\"odinger-type 
 differential equation
\begin{equation}
-\frac{d^{2}{\bar \Delta}}{dx^{2}}+V\left(x\right){\bar \Delta}=0,
\label{eq:SE_eps}
\end{equation}
with potential
\begin{equation}
V\left(x\right)=\left(1+\epsilon\right)\left(1+\epsilon-4g K\left(p_{0}e^{x}\right)\right).
\label{eq:SE_epsV}
\end{equation}
The boundary conditions are now given as
\begin{eqnarray}
{\bar \Delta}'\left(0\right) & = & \left(1+\epsilon\right){\bar \Delta}\left(0\right), \nonumber \\
{\bar \Delta}'\left(Y\right) & = & -\left(1+\epsilon\right){\bar \Delta}\left(Y\right),
\end{eqnarray}
where
 $Y=\log\frac{\Lambda}{p_{0}} 
 = \log{\frac{A \Lambda^{2\epsilon}}{T}}$
  and $T = T_c (\epsilon)$.
  Below, we solve the gap equations in the  limits 
where the ratio of the small dimensionless
constants $g$ and $\epsilon$ is either large or small. 
In each case we 
compare the analytical results with the numerical solution of the  gap equation.

\subsubsection{The limit $\epsilon \ll g$}
\begin{figure}
    \centering
    \includegraphics[width=0.8\linewidth]{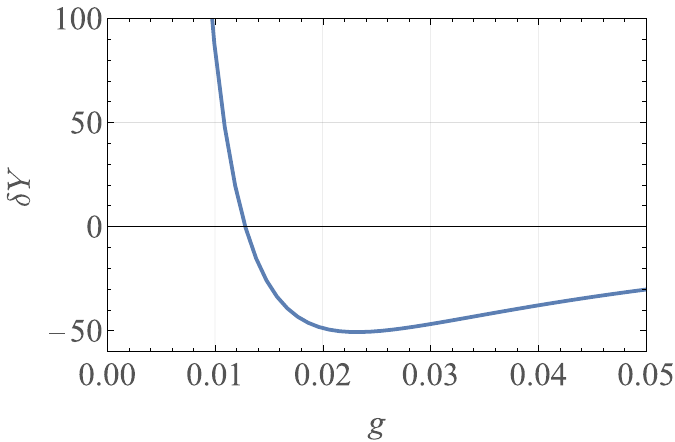}
    \caption{The 
     dependence of $\delta Y$ from Eq.~(\ref{yy_999}) on the dimensionless coupling $g$, obtained 
      from the numerical solution of the gap equation at an extended VH point, expanded to first order in $\epsilon$.   
      The sign of $\delta Y$ determines whether $T_c$ for an extended VH point increases or decreases with 
     $\epsilon$.  For a negative $\delta Y$, $T_c$ increases, for a positive $\delta Y$ it decreases.  
We see that $\delta Y$ is positive at very small small $g$ and negative at larger $g$. The sign change occurs at  $g^*\approx 0.013$. For small $g$, $\delta Y$ scales as $1/g^2$. }
    \label{fig:small_epsilon_correction}
\end{figure}
  In this limit, we compute the leading correction in $\epsilon/g$ to the expression for $T_c$ for an ordinary VH point from
first order perturbation theory for the 
Schr\"odinger equation, expanding the
equation, the boundary condition, and the
potential $V\left(x\right)$ to linear order in $\epsilon$. The resulting
set of equations is then solved numerically and determines the correction $\delta Y$
defined as
\begin{equation}
Y (\epsilon)= 
Y(\epsilon=0) +\epsilon\delta Y.
\label{yy_999}
\end{equation}
The result for $\delta Y$
is shown in Fig.~\ref{fig:small_epsilon_correction}. 
When $g$ is larger than $g^{*}\approx0.013$, 
$\delta Y$ is negative, hence the superconducting transition temperature
 increases with $\epsilon$. 
At smaller
$g<g^{*}$, 
$\delta Y>0$ and scales with $g$ as $1/g^2$.
The superconducting transition temperature decreases with $\epsilon$ as
\begin{equation}
T_c (\epsilon) = T_c (\epsilon=0) e^{-a \epsilon/g^2}
\end{equation}
where $a = O(1)$.
We emphasize that this $T_c (\epsilon)$
smoothly connects to 
 $T_c (\epsilon =0)$ 
 at the ordinary VH point. 

\subsubsection{The limit $\epsilon \gg g$}

We now show how the result for $T_c$ gets modified in the opposite limit $\epsilon \gg g$. We set $\epsilon$ to be a number of order one and  compute $T_c$ by order of magnitude, keeping the explicit dependence on $\epsilon$ in the exponent,
 but neglecting the dependence on $\epsilon$ in the prefactor.
 For $\epsilon \gg g$,  we can safely take the limit $\Lambda \rightarrow \infty$ as all integrals are UV convergent. 
  Evaluating the integral for $K(p_+)$ in (\ref{yy_9999}) in infinite limits, we obtain 
\bea
 \lambda_\epsilon &\equiv&  2 g K(p_+)
 \nonumber \\
&=& \frac{4\epsilon^{2}}{\log2}\int_{x_{\rm min}}^{\infty}\frac{dx}{|1+x|^{2(1+\epsilon)}-|1-x|^{2(1+\epsilon)}},
\label{w_8}
\eea
 where 
 $x = p_-/p_+$ and 
 $x_{\rm min} \sim T/T_\epsilon$, where $T_\epsilon =A (p_+)^{2(1+\epsilon)}$. 
  Next, we 
   assumed and verified that the relevant $p_+$ in the equation for the pairing vertex 
    are of order $\Lambda g^{1/(4\epsilon)}$.
  This
    allows us to express 
    $T_{\epsilon} \sim A\Lambda^{2\left(1+\epsilon\right)}  g^{(1+\epsilon)/(2 \epsilon)}$.

The integral over $x$ 
in (\ref{w_8})
converges in the UV limit, which allows us to set the upper limit of the integration over $x$ to infinity. It is logarithmically singular in the IR limit and with logarithmic accuracy we  obtain
$\lambda_\epsilon \sim \log{T_\epsilon/T} 
= \log{T_\epsilon/T_c (\epsilon)}  $. 
Returning  to the gap equation \eqref{eq:SE_eps},  we now have 
 $V(x) \approx (1+ \epsilon) (1 + \epsilon -2 \lambda_\epsilon) = \beta^2_\epsilon$.
 The solution of Eq.~\eqref{eq:SE_eps} with such $V$ is 
${\bar \Delta}\left(x\right)\sim e^{\pm\beta_\epsilon x}$ 

\begin{figure}
    \centering
    \includegraphics[width=\columnwidth]{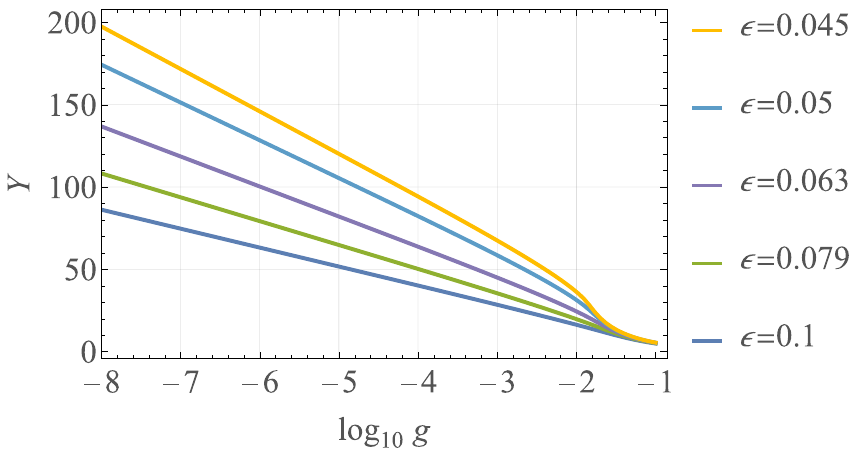}
    \caption{Dependence of $Y\left(g,\epsilon\right)=\log\frac{T_{0}}{T_{c} (\epsilon)}$
on the coupling constant $g$,
 obtained by solving the gap equation numerically (see Sec. \ref{sec:num1}). For $\epsilon>g$ we find power-law
behavior $T_{c}\propto g^{\alpha}$. 
The exponent $\alpha$ is determined 
from the slope of $Y\left(g,\epsilon\right)$ vs $\log g$ 
 and with high accuracy is  $\alpha\approx\frac{1}{2\epsilon}$, consistent with our analytical analysis.
For $\epsilon<g$ the behavior deviates from the 
power-law dependence.
    }
    \label{fig:TceVHnum}
\end{figure}

A similar power-law solution
 (as a function of frequency) holds for a number of quantum-critical systems~\cite{Chubukov_gamma}
 and Yukawa SYK-type models~\cite{Schmalian_syk,Wang_syk,Classen_syk}.  We verified that, like there,
 the solution, that satisfies boundary conditions, does not exist when $\beta_\epsilon$ is real, but emerges when $\beta_\epsilon$  becomes complex, and the onset of complex $\beta_\epsilon$ sets  the value of $T_c$.  In our case, $\beta_\epsilon$  becomes complex at  $\lambda_\epsilon = (1+\epsilon)/2$, which for a generic $\epsilon$ is a number of order one.   Using 
 $\lambda_\epsilon \sim \log{T_\epsilon/
 T_c (\epsilon)}$, 
 we then find that
 \beq
 T_c (\epsilon) \sim T_\epsilon \sim T_0 g^{({1+\epsilon})/({2 \epsilon})},
 \label{w_11}
 \eeq
 where $T_0 \sim A \Lambda^{2(1+ \epsilon)}$.
 We see that $T_c (\epsilon)$ is 
 not exponentially small in $g$.
 We also notice that
  Eq.~(\ref{w_11}) can be re-expressed,
   using (\ref{ee_16}), 
   as  
\begin{equation}
T_{{\rm c}} (\epsilon) \sim A^{-1/\epsilon}U^{\frac{1+\epsilon}{\epsilon}}.
\end{equation}
This last expression shows that $T_{{\rm c}} (\epsilon)$ does not depend on the upper cutoff $\Lambda$ (the $\Lambda-$dependencies in $T_0$ and $g$ cancel out) and in this respect is universal. 

 At a qualitative level, we found that 
 the crossover from $T_c$ for an ordinary VH point to the one for an extended VH point 
is captured by the interpolation formula 
 \begin{equation}
\frac{1+\epsilon}{2\epsilon}\left(\left(\frac{T_{0}}{T_{c} (\epsilon)}\right)^{\frac{2\epsilon}{1+\epsilon}}-1\right) =\frac{1}{\gamma g}.
\end{equation}
In the limit $\epsilon \ll g$, this reduces to $\log{T_0/T_c (\epsilon \to 0)} = 1/\gamma g$,  in the opposite limit $\epsilon \gg g$, one recovers 
the universal power-law expression $T_c (\epsilon) \sim T_0 g^{(1+ \epsilon)/(2\epsilon)}$.

\begin{figure}
    \centering
    \includegraphics[width=\columnwidth]{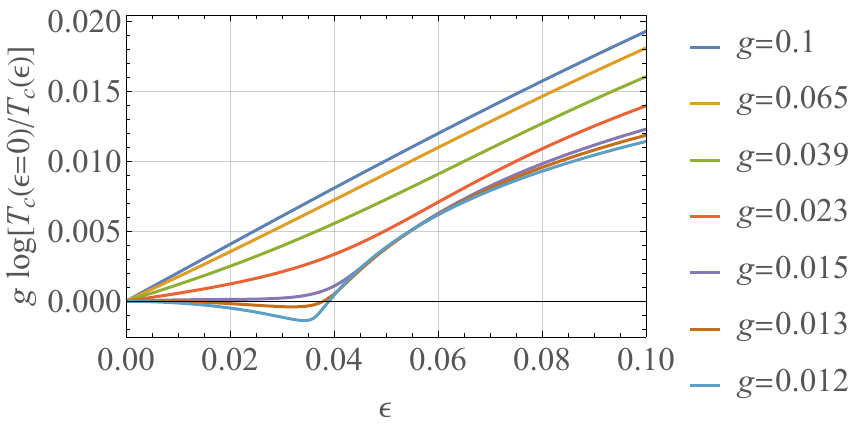}
    \caption{Dependence of $g \log\left[T_c (\epsilon=0)/T_c (\epsilon)\right]$
    on $\epsilon$ for different values of  $g$,
    obtained by solving the gap equation numerically  (see Sec. \ref{sec:num1}). 
    For $g < g^* = 0.013$, the dependence is non-monotonic, consistent with our analytical results.
    For larger $g$, $T_c (\epsilon)$ monotonically increases with $\epsilon$.}
    \label{fig:nonmonotonic_epsilon}
\end{figure}

\subsubsection{Numerical solution for extended saddle points}
\label{sec:num1}
  We also 
  performed the integration 
   over the transverse momenta in the function $K(p_+)$  in (\ref{yy_9999}) numerically and solved numerically the gap equation (\ref{yy_99}) with this $K(p_+)$. 
  We show the results in Figs. ~\ref{fig:TceVHnum} and \ref{fig:nonmonotonic_epsilon}.
   In Fig.~\ref{fig:TceVHnum} we show the dependence of
   $Y(g,\epsilon)=\log\left(\frac{T_0}{T_c (\epsilon)}\right)$
on the coupling constant $g$. The distinct behavior for $\epsilon$ smaller and larger $g$ is clearly visible.  For $\epsilon>g$  the linear dependence of $Y$
 on 
 $\log g$ demonstrates that the transition temperature $T_c (\epsilon)$ 
 has the 
power law form $T_c (\epsilon) \propto g^\alpha$.
The  value of $\alpha$ is determined from the slope. The data are  
 best described by 
$T_{c} (\epsilon) \propto g^{\frac{1}{2 \epsilon}}$,  consistent with  (\ref{w_11}).
For $\epsilon<g$ the behavior deviates from the power law. In Fig.~\ref{fig:nonmonotonic_epsilon} we show how the small $\epsilon$ behavior of $T_c (\epsilon)$  from 
Fig.~\ref{fig:small_epsilon_correction} interpolates to the power-law behavior at larger $\epsilon$.
For $g<g^* =0.013$, the dependence of $T_c (\epsilon)$ on $\epsilon$ is non-monotonic, 
in agreement with our analytic findings.

\section{Pairing fluctuations, BKT transition and charge-$4e$ superconductivity}

We next discuss the statistical mechanics that we expect
to emerge from our analysis. 

The solutions discussed in the previous section formally belong to
a two-dimensional irreducible representation $\left(\Delta\left(k_{+}\right),\Delta\left(k_{-}\right)\right)$
of the point group. Hence, fluctuations of the order parameter might
give rise to vestigial order with symmetry breaking of composite
order parameters like nematic or time-reversal symmetry breaking states\cite{Fernandes2019}. This
is a consequence of the four-fold symmetric dispersions of Eqs.~\eqref{eq:VHdisp} and \eqref{eq:extVH}.
However, the symmetry of a single VH point is usually lower and Eqs.~\eqref{eq:VHdisp} and \eqref{eq:extVH} 
are the result of an anisotropic rescaling of momenta.
This will lift the degeneracy of the two solutions and the triplet order parameter  belongs to a one-dimensional representation of the point group.

For a 
2D
system one usually expects a Berezinskii--Kosterlitz--Thouless (BKT) transition\cite{Kosterlitz1973,Kosterlitz1974}
to a state with algebraic order and with finite superfluid stiffness.
As discussed by Halperin and Nelson\cite{Halperin1979}, the resulting BKT transition
temperature is 
very close to the mean field transition temperature
that one obtains from the solution of 
 the 
 gap equation.
The reason is that the threshold stiffness of the BKT transition is
much smaller than the 
low-$T$ stiffness of a 
 weakly coupled 
 superconductor. Hence vortex proliferation sets in only very near the
mean field transition temperature. However, the BKT
 physics does not hold for a 
 triplet superconductor 
without spin-orbit interaction. 
 Fluctuations of such a state 
  are governed by the
three-component complex 
 coordinate-dependent field 
 $\boldsymbol{\psi}
 (x) =\left(\psi_{1} (x),\psi_{2} (x),\psi_{3} (x)\right)^{T}$
that describes 
long-wavelength variations of the pairing wave
function
\begin{eqnarray}
\Psi_{\alpha\beta}\left(\boldsymbol{k},\boldsymbol{x}\right) & = & \sum_{i=1}^{3}\psi_{i}\left(\boldsymbol{x}\right)\Delta_{i}\left(\boldsymbol{k}\right)\left(\sigma^{i}i\sigma^{y}\right)_{\alpha\beta}
\end{eqnarray}
where $\Delta_{i}\left(\boldsymbol{k}\right)$ is the gap function
discussed in the previous section. 
Fluctuations between components  
of $\boldsymbol{\psi}$,
i.e. fluctuations in the spin sector of the triplet state,
destroy
even 
an
algebraic order due to the Hohenberg--Mermin--Wagner theorem\cite{Hohenberg1967,Mermin1966}.
In the notation $\boldsymbol{\psi}=\psi_{0}\boldsymbol{n}e^{i\theta}$
where $\theta$ is the $U(1)$ phase of the superconductor while the unit vector $\boldsymbol{n}$ describes spin fluctuations of the triplet state, algebraic order of $\boldsymbol{\psi}$ is  suppressed by fluctuations of $\boldsymbol{n}$. 

The situation 
 is different for a 
 composite order parameter
\begin{equation}
\phi\left(\boldsymbol{x}\right)=\boldsymbol{\psi}\left(\boldsymbol{x}\right)\cdot\boldsymbol{\psi}\left(\boldsymbol{x}\right),
\end{equation}
which describes a  charge-$4e$ superconductor,
 in which
two triplets form a singlet in spin space\cite{Korshunov1985,Schmalian2021}.
Since $\mathbf{n}^2=1$
it follows that
\begin{equation}
\left\langle \phi^{*}\left(x\right)\phi\left(x'\right)\right\rangle =\psi_{0}^{4}\left\langle e^{-2i\left(\theta\left(x\right)-\theta\left(x'\right)\right)}\right\rangle,
\label{ee_18}
\end{equation}
i.e., $\phi$ possesses only  phase fluctuations, which allow  a BKT transition.  The extra factor $2$ in the exponent in 
(\ref{ee_18}) 
allows for fractionalized
vortices of the primary superconducting order parameter 
(the spin field  heals  the mismatch that forms
at a fractional vortex, see Ref.~\cite{Mukerjee2006}).
 The threshold stiffness  for the BKT transition in a charge-$4e$ superconductor 
is four times larger than
that for a charge-$2e$ superconductor,  yet it is still much smaller than the
zero temperature stiffness. Hence,   the  $4e$ BKT transition still occurs e
very near  mean-field $T_{c}$ for the primary $2e$ order parameter. 
If  spin--orbit interaction is present,
the anisotropy in spin space  suppresses
fluctuations and allows  charge-$2e$ superconductivity  with an algebraic  order.

\section{Summary}

In this work we 
analyzed
low-temperature instabilities of 
 a system of fermions  
 with 
 a
 single ordinary or extended 
  VH point at the Fermi level, 
  in the limit of small electron-electron interactions. 

We first considered the possibility of ferromagnetic order and argued that it likely does not develop because of strong reduction of   
particle-hole response in the  $\mathbf{q}\rightarrow \mathbf{0}$ limit by 
  particle-particle fluctuations.
 
We then analyzed  pairing instabilities and explicitly demonstrated both, analytically and numerically, 
 that a system with a   single  VH point at the Fermi level is unstable towards  triplet superconductivity.  The instability develops for both a conventional and a higher-order VH point, but $T_c$ is much higher for a higher-order VH point  in the regime $\epsilon > g$, where it varies with the coupling constant $g$  in a power-law fashion, as $T_c \propto g^{(1+ \epsilon)/(2\epsilon)}$.
 We showed that this $T_c$ is a universal, cutoff independent quantity, determined by  the  band curvature 
 and  the  local interaction.

The  attractive triplet component of the pairing vertex   comes from the Kohn-Luttinger type dressing of the pairing interaction by 
particle-hole fluctuations. 
Yet, we  demonstrated that the mechanism for superconductivity is distinct from the usual Kohn-Luttinger one.
In our problem, the attractive component of the vertex function 
is weak and, on its own, would not  lead to a Cooper instability. 
However,  the enhancement of the  density of states  near a VH point   overcomes the smallness  of the pairing vertex and gives 
 rise to a BCS-like expression for $T_c$ for an ordinary VH point and to 
 power-law dependence of $T_c$ on the coupling   for an extended VH point.
 
 As a consequence of this fundamentally non-BCS pairing mechanism, 
 the transition temperature is expected to rapidly drop once  
 the system moves away from a VH singularity under, e.g., a  change of  the chemical potential.
  Given the absence of a Cooper phenomenon, we 
  expect that away from a VH singularity, superconductivity will develop only if the 
   coupling exceeds a certain threshold. For any given $g$, there will be then 
   a 
   superconducting quantum critical point at some detuning.  Similar behavior occurs  for  pairing at a critical point towards density-wave order and in SYK models~\cite{Chubukov_gamma,Schmalian_syk,Wang_syk,Classen_syk, Chubukov_gamma_a}. 
   \begin{figure}
    \centering
    \includegraphics[width=\columnwidth]{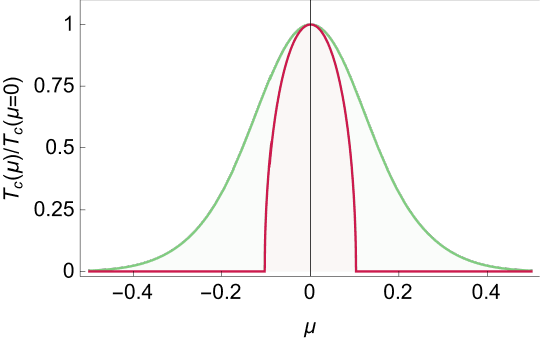}
    \caption{
    A schematic plot of the dependence of $T_c$ on  detuning from the VH point by a parameter $\mu$ (e.g. 
     a non-zero chemical potential)
     In our 
      case (red curve), the transition temperature, determined 
       by Eq.~\eqref{eq:detB}, vanishes when 
       detuning  exceeds $T_c(\mu=0)$. 
        In the case of a  constant attraction (green curve), the transition temperature, determined by 
        Eq.~\eqref{eq:detA}, gets reduced upon detuning from the VH point, but remains finite.}
    \label{fig:de-tuning}
\end{figure}

   To illustrate this effect we compare $T_c (\mu)$ for our system and for a system with a constant attraction $g$ and a 
   logarithmic density of states, detuned 
   by $\mu$ (a detuned version of the model discussed by Son, see \ref{Sonsmodel}).
 In the last case, the transition temperature is determined
by
\begin{equation}
1=\frac{4}{\pi^{2}}g\log\left(\frac{T_{0}}{T_{c}}\right)\log\left(\frac{T_{0}}{\sqrt{T_{c}^{2}+\mu^{2}}}\right),
\label{eq:detA}
\end{equation}
where $T_c (\mu =0) \propto e^{-\pi/(2 \sqrt{g})}$, see Eq.~\eqref{ch_18}. 
For $\left|\mu\right|>T_{c} \left(\mu=0\right)$,  
the effective coupling
constant is reduced, yet 
$T_{c} (\mu)$ remains finite.
For our problem, 
$T_c (\mu)$ is determined by 
\begin{equation}
1=\gamma g\log\left(\frac{T_{0}}{\sqrt{T_{c}^{2}+\mu^{2}}}\right)
\label{eq:detB}
\end{equation}
and remains non-zero 
only at 
$\left|\mu\right|<T_{c}\left(\mu=0\right)$.
For larger detuning from a VH point, $T_{c}=0$.
This sets a 
 superconducting quantum critical point at $|\mu| \sim T_c (\mu =0)$.
We illustrate this in Fig.~\ref{fig:de-tuning}.

 We also discussed the role of critical fluctuations and 
  argued that the transition temperature that we derived from the linearized gap equation
   is close to  
   a BKT transition into an algebraic superconductor, which in the absence of spin-orbit interaction is  a charge-$4e$ superconductor made up of singlet bound states of triplet pairs, and in the presence of spin-orbit interaction is a 
   charge-$2e$ superconductor.
  
  In our  analysis we concentrate on processes that are exclusively due to interactions
   between fermions at or near 
    a 
     VH point. It is important to keep in mind  that  
    some
    crucial physical processes may 
     come from 
     electronic states away from a VH point, 
      particularly 
      for 
transport phenomena\cite{Stangier2022}. For thermodynamic instabilities,  the   
instability that we found here is, however, the leading  one in the pairing channel in the limit of weak coupling.

\begin{acknowledgments}
We are grateful to Erez Berg, Joseph Betouras,
Anzumaan Chakraborty,
Laura Classen, Elio König, Mohid Randeria, Michael Scherrer,
and Veronika Stangier for helpful discussions. This work was supported
by the German Research Foundation TRR 288-422213477 ELASTO-Q-MAT,
Projects A11 (MG) and B01 (RO, JS), a Weston Visiting Professorship
at the Weizmann Institute of Science (JS), and by 
National Science Foundation grant NSF: DMR - 2325357 (AVC and YCL).
Part of the work was performed when 
JS and AVC visited 
KITP at UCSB.  KITP is supported in part
by the National Science Foundation under PHY-1748958.
\end{acknowledgments}

\appendix
\widetext
\setcounter{equation}{0}
\renewcommand{\theequation}{S\arabic{equation}}
\makeatletter 
\setcounter{figure}{0}
\renewcommand{\thefigure}{S\@arabic\c@figure}
\makeatother

\section*{Appendix A. A potential ferromagnetic instability}

 In this Appendix we present the details of our analysis of a potential instability towards ferromagnetism.
 For definiteness, we consider a single ordinary VH point with dispersion $\epsilon_k = (k^2_x - k^2_y)/(2m)$.
  The Stoner instability can be detected by computing the static and uniform magnetic susceptibility $\chi$ (the limit $q\to 0$ of the static $\chi (q)$). This susceptibility can be obtained by either introducing an infinitesimal magnetic field and computing magnetization or by introducing an infinitesimal bare ferromagnetic order parameter, dressing it by the interactions,  and computing the ratio of the fully dressed and the bare order parameters. In the diagrammatic analysis, the second approach is easier to implement. The bare order parameter  $\Delta_0$ is represented as a two-particle vertex, and the  dressed one $\Delta$ is obtained by renormalizing this vertex  by the interactions (see Fig.~\ref{fig:Stoner}(c)).  
  In the  ladder approximation,
  $\Delta = \Delta_0 \left(1 + U \Pi_{\rm ph} (0,0) + (U\Pi_{\rm ph} (0,0))^2 + ...\right)$, where $\Pi_{\rm ph} (0,0)$ is the static
   particle-hole bubble $\Pi_{\rm ph} (q,\Omega =0)$ in the limit $q=0$.  The perturbative series are geometrical, and
   the  susceptibility, defined as $\chi  =\Delta/\Delta_0$, is
   \beq
   \chi =\frac{1}{1 - U \Pi_{\rm ph} (0,0)}.
   \label{ap_1}
   \eeq
   For an ordinary VH point,
  \begin{equation}
  \Pi_{\rm ph} (0,0) = \frac{m l}{2\pi^2},
  \end{equation}
  where
  \beq
  l=\log{\frac{\Lambda^2}{mT}}
  \eeq
  and  $\Lambda \sim p_F$ is the upper momentum cutoff.
 Within this approximation, a ferromagnetic instability develops at any $U$, at a temperature much larger than the superconducting $T_c$, which we obtained in the main text.  If this was the case, our consideration of the pairing instability would be invalid.

The summation of the ladder series can be reformulated in the RG language as the solution of the differential RG equation
 for the running $\Delta (l)$ in terms of the running static coupling $\lambda_{\rm ph} (l)$ at zero momentum transfer.
 The equations for running $\Delta (l)$ and $\lambda_{\rm ph} (l)$ are
 \bea
  &&  \frac{d \Delta (l)}{dl} = \lambda_{\rm ph} (l) \Delta (l),  \label{ap_2}\\
 && \frac{d \lambda_{\rm ph} (l)}{dl} = \left(\lambda_{\rm ph} (l)\right)^2,
 \label{ap_3}
  \eea
and the bare value of the coupling is
  \begin{equation}
\lambda_{\rm ph} (0)= \lambda_0 =\frac{mU}{2\pi^2}.
  \end{equation}
  The solution of these equations is Eq.~(\ref{ap_1}) with $U \Pi_{\rm ph} (0,0) = \lambda_0 l$, i.e.,
  \beq
   \chi =\frac{1}{1 - \lambda_0 l}.
   \label{ap_1_1}
\eeq
  As we said in the main text, the ladder approximation should not be trusted in our case because
   the two-particle vertex in Fig.~\ref{fig:Stoner} also get renormalizations from the particle-particle channel.
    The static and uniform $\Pi_{\rm pp} (0,0)$ scales as $l^2$ due to the combination of the logarithmic singularity in the density of states and the Cooper logarithm:
     \beq
     \Pi_{\rm pp} (0,0) = -\frac{m}{4\pi^2} l^2.
     \eeq
The polarization $\Pi_{\rm pp} (0,0)$ renormalizes the static coupling $\lambda_{\rm pp} (l)$ with zero total incoming momentum.
 The  ladder series of the renormalizations in the particle-particle channel yield
 \be
 \lambda_{\rm pp} (l) = \frac{\lambda_0}{1 + \lambda_0 l^2}.
 \ee
 Because $\lambda_0 >0$, the running coupling in the particle-particle channel decreases as $l$ increases.
 This behavior can again be reformulated in RG as the flow equation
 \be
 \frac{d \lambda_{\rm pp} (l)}{dl} = -2 l \left(\lambda_{\rm pp} (l)\right)^2.
  \ee
  Beyond the ladder approximation, there are cross-renormalizations of $\lambda_{\rm ph} (l)$ in the particle-particle channel and of $\lambda_{\rm pp} (l)$ in the particle-hole channel. Within RG, it seems natural to add the two contributions to the RG flow of each coupling. For $\lambda_{\rm ph} (l)$, this would mean that the RG equation becomes, instead of (\ref{ap_3}),
  \be
 \frac{d \lambda_{\rm ph} (l)}{dl} = \left(\lambda_{\rm ph} (l)\right)^2 \left(1-2l\right).
 \label{ap_4}
  \ee
 Solving (\ref{ap_4}) one would then obtain
  \be
  \lambda_{\rm ph} (l) = \frac{\lambda_0}{ 1 + \lambda_0 l \left(l-1\right)}
 \label{ap_5}
  \ee
  and
   \be
 \chi (l) =  \exp\left[\left(\frac{2\lambda_0}{4-\lambda_0}\right)^{1/2} \left(\arctan{(2l-1)\left(\frac{\lambda_0}{4-\lambda_0}\right)^{1/2}} + \arctan{\left(\frac{\lambda_0}{4-\lambda_0}\right)^{1/2}}\right)\right].
 \label{ap_6}
  \ee
 This $\chi(l)$ saturates at large $l$ instead of diverging, hence within RG there is no ferromagnetic instability: renormalizations in the particle-hole channel, which  increase the running couping and would nominally lead to such an instability, are overshoot by stronger
   renormalizations in the particle-particle channel, which reduce the coupling.

   This analysis, however, assumes that the renormalizations of the coupling in the two channels just add up.
   This needs to be verified because the $l^2$ renormalization in the particle-particle channel holds for the interaction
   $\lambda_{\rm pp}$ with zero total momentum, while for the analysis of potential  ferromagnetism  we need to know $\lambda_{\rm ph}$ at zero momentum transfer.   For this reason, below we explicitly compute the two-loop diagram for the renormalization of the vertex $\Delta (l)$, by combining the renormalizations in the particle-hole and particle-particle channels.  We show the corresponding diagram in Fig.~\ref{fig:Stoner}(c).  The one-loop vertex renormalization holds in the particle-hole channels and yields the correction $\lambda_0 l$, If the RG treatment is correct, at least qualitatively,  the mixed two-loop  diagram must give $O(l^3)$.   We show below that the $l^3$ term vanishes and the actual dependence is $l^2$.

   The calculation proceeds in a standard way.  We set external momenta to values at the VH point and the external fermionic Matsubara frequency to $\pi T$, and use the temperature as IR cutoff.  The quantity we need to calculate is
   \be
   X = \int_p G^2 (p-p_0) \Pi_{\rm pp} (p),
   \ee
    where $p =({\bf p}, \Omega)$ and $\Omega$ is Matsubara frequency, and $p_0 = (0, \pi T)$.
   We use momentum variables $p_{+-} = (p_x \pm p_y)/\sqrt{2}$. In these variables, the fermionic dispersion is 
   $\epsilon_ {\bf p} = p_+ p_- /m$.

   The polarization bubble $\Pi_{\rm pp} (p)$ is the convolution of the Green's functions of two fermions with   $l = ({\bf l}, \omega)$ and $l+p = ({\bf l} + {\bf p}, \omega + \Omega )$. Summing over  $\omega$, we obtain
\be
    \Pi_{\rm pp} (p) = \int \int \frac{dl_+ dl_-}{4\pi^2} \frac{1- n_F \left(\frac{l_+ l_-}{m}\right) - n_F \left(\frac{(l_++p_+)(l_- +p_-)}{m}\right)}{\epsilon_{l,p} -i\Omega}
   \ee
   where 
   $\epsilon_{l,p} = \frac{l_+ l_-}{m} + \frac{(l_++p_+)(l_- +p_-)}{m}$.
 Combining this with $G^2 (p-p_0)$, summing over $\Omega$, and neglecting $\pi T$, which is irrelevant to the analysis of the power of $l$, we obtain $X = X_1 + X_2$, where
\be 
X_1 = \int \int \frac{dl_+ dl_-}{4\pi^2}  \int \int \frac{dp_+ dp_-}{4\pi^2} \left(1- n_F \left(\frac{l_+ l_-}{m}\right) - n_F \left(\frac{(l_++p_+)(l_- +p_-)}{m}\right) \right) \frac{n_F (\epsilon_{l,p}) - n_F (\epsilon_p)}{(\epsilon_p-\epsilon_{l,p})^2}
   \ee
  and
  \be
  X_2 = \int \int \frac{dl_+ dl_-}{4\pi^2}  \int \int \frac{dp_+ dp_-}{4\pi^2} \frac{1}{4 T \cosh^2{\frac{\epsilon_p}{2T}}} \frac{1- n_F \left(\frac{l_+ l_-}{m}\right) - n_F \left(\frac{(l_++p_+)(l_- +p_-)}{m}\right)}{\epsilon_{l,p} - \epsilon_p}.
   \ee
    The upper limit of momentum integration is $\Lambda$, the lower limit is effectively set by $(mT)^{1/2}$  because of Fermi functions.

  Each of the two terms gives $O(l^3)$.  This can be seen most straightforwardly by evaluating $X_2$.
   Here, typical $\epsilon_p$ are of order $T$.  Re-expressing $\int \int dp_+ dp_- /(4\pi^2)$ as
   $(m/4\pi^2) \int dp_+/p_+ \int d \epsilon_p$ and keeping typical $l_{+}, l_-$, and $p_+$ much larger than $T$ by absolute value, in anticipation of the logarithms, one can re-express $X_2$ as
   \be
   X_2 = \frac{m}{16 \pi^4} \int \frac{dp_+}{|p_+|} \int \int dl_+ dl_-\frac{1- n_F \left(\frac{l_+ l_-}{m}\right) - n_F \left(\frac{(l_++p_+)(l_- +p_-)}{m}\right)}{\epsilon_{l,p}} \times I
  \label{ap_7}
  \ee
   where the integral $I$ simplifies due to
   \be
   I = \frac{1}{4T} \int_{-\infty}^{\infty} \frac{d \epsilon_p}{\cosh^2 {\frac{\epsilon_p}{2T}}} =1.
   \ee
   To estimate the $l-$dependence of $X_2$, we note that for $l >p$ and $\epsilon_l > mT$,  the numerator in (\ref{ap_7}) is $\text{sign}{\epsilon_l}$ and $\epsilon_{l,p} \approx \epsilon_l = l_+ l_-/m$.
    Rescaling then $p_+$, $l_+$ and $l_-$ by $(mT)^{1/2}$, converting the integration to positive variables, and using the squares of the original variables as the new ones, which we label $a, b$ and $c$,   we obtain
    \be
  X_2 = \frac{m^2}{16 \pi^4} \int_1^l \frac{da}{a} \int_1^l \frac{db}{b} \int_1^l\frac{dc}{c} = \frac{m^2}{16 \pi^4}  l^3.
  \label{ap_8}
  \ee
   A similar analysis can be done for  $X_1$. Here we note that
   \be
   \int d \epsilon_p \frac{n_F (\epsilon_{l,p}) - n_F (\epsilon_p)}{(\epsilon_p-\epsilon_{l,p})^2} \approx
   - \frac{1}{\epsilon_{l,p}}
   \ee
    and hence
    \be 
    X_1 =- \frac{m}{16 \pi^4} \int \frac{dp_+}{|p_+|} \int \int dl_+ dl_-\frac{1- n_F \left(\frac{l_+ l_-}{m}\right) - n_F \left(\frac{(l_++p_+)(l_- +p_-)}{m}\right)}{\epsilon_{l,p}} \times I.
  \label{ap_9}
  \ee
  Evaluating the remaining integral in the same way as we did for $X_2$, we find $X_1 = - \frac{m^2}{16 \pi^4}  l^3$.
  However, comparing (\ref{ap_9}) and (\ref{ap_7}), we see that they are exactly opposite to each other.  This holds even before we approximate each term by (\ref{ap_8}). As a result, the $l^3$ terms in $X_1$ and $X_2$ cancel each other, even if we compute each with more care than we did in moving from (\ref{ap_7}) to (\ref{ap_8}).
  We emphasize that to detect the cancellation, one must keep frequency dependence in $\Pi_{\rm pp} (p)$.  If we approximated
  the particle-particle polarization by its static form $\Pi_{\rm pp} ({\bf p}, \Omega =0)$, we would get  $X = X_2$, with no $X_1$ term.  In this situation, the $l^3$ term in $X$ would be present.

  The next term in $X$ is of order $l^2$, and there is no reason why such a term may cancel.  We computed $X$ numerically
  and did find that $X$ scales as $l^2$.   Explicitly, we found
  \be
  X \approx - b\left(\lambda_0 l\right)^2,
  \ee
  where $b \approx 1.8$.
  To order $\lambda^2_0$, the renormalization in the particle-particle channel then changes the bare coupling to
  $\lambda_{\rm eff} = \lambda_0 (1 - b \lambda_0 l)$.  We didn't compute higher-order terms, but it is reasonable to assume that the series are geometrical, at least approximately, in which case $\lambda_{\rm eff}$ can be approximated by $\lambda_{\rm eff} = \lambda_0/(1 + b \lambda_0 l)$.  
  Using then this $\lambda_{\rm eff}$ instead of $\lambda_0$ in (\ref{ap_1_1}), we find
  \beq
   \chi =\frac{1}{1 - \lambda_{\rm eff}l} \propto \frac{1}{1 - \lambda_0 l (1-b)}.
   \label{ap_1_2}
   \eeq
   The same result for $\chi$ is obtained if we replace $1-2l$ in the r.h.s. of (\ref{ap_4}) by $1-b$.
  The outcome then depends on the magnitude of $b$. For $b >1$, consistent with our numerical result, a ferromagnetic
  susceptibility decreases as $l$ increases. Then a ferromagnetic instability does not develop.

\begin{figure}
      \centering
      \includegraphics[width=0.9\columnwidth]{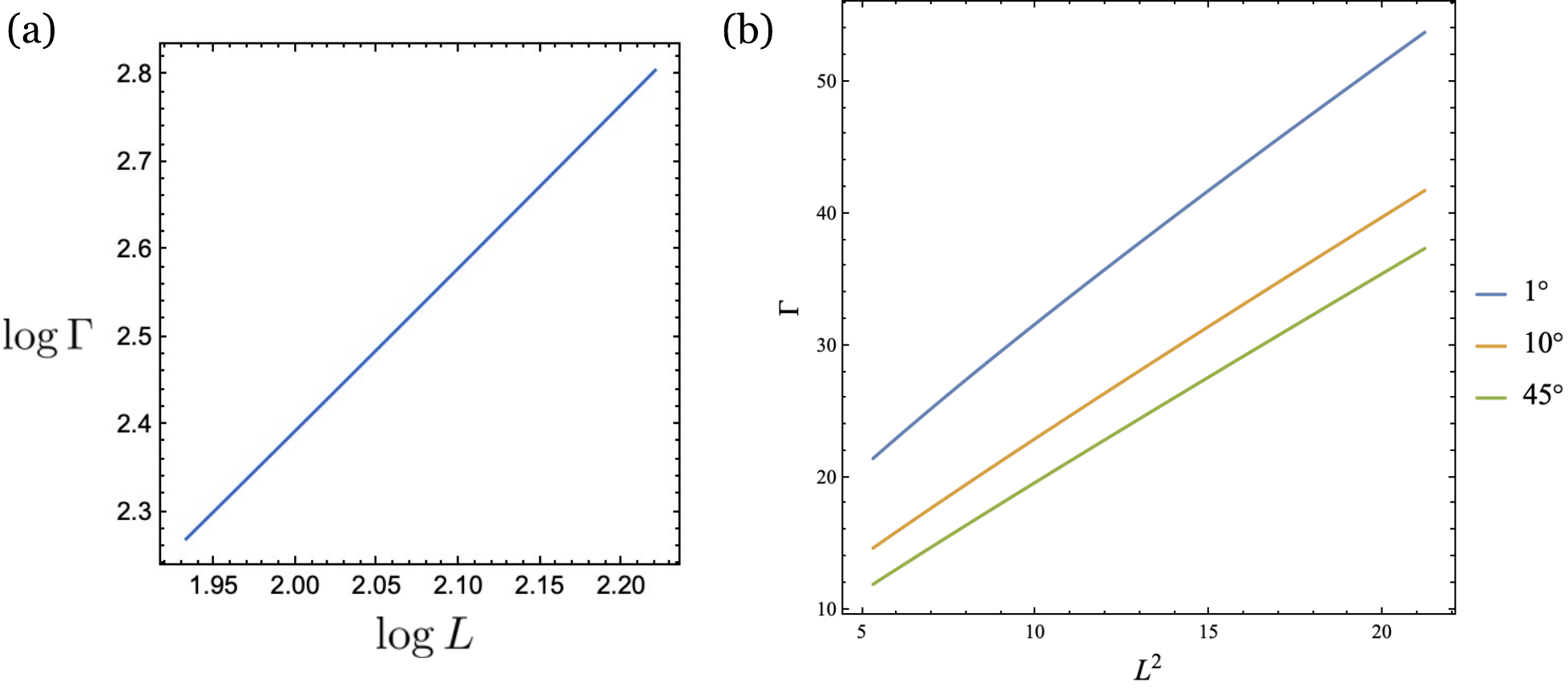}
      \caption{(a) The log-log plot of $\Gamma$, given by Eq.~(\ref{y_888}) vs $L=\log{\frac{\Lambda}{|\textbf{p}|}}$. The slope gives the value of the exponent of the power-law dependence of $\Gamma$ on  $L$.  To a good accuracy, $\Gamma = b' L^2$. 
      (b) $\Gamma$ as a function of $L^2$ for various polar angles of $\textbf{p}$ relative to the $k_+-$direction. For all angles, $b'$ is in the range between $1.4$ and $1.9$.}
      \label{fig:Gamma}
  \end{figure}

  The two-loop diagram in Fig.~\ref{fig:Stoner}(c) can also be evaluated at zero temperature and regularized by a small but finite external momentum \textbf{q}. In this case, the quantity we are computing is
\begin{equation}
\begin{aligned}
    \Gamma=\int\int \frac{d\omega d\Omega}{4\pi^2} \int\int\frac{dk_+dk_-}{4\pi^2}\int\int\frac{dq_+dq_-}{4\pi^2}G(q+p)^2G(-k+q/2+p)G(k+q/2+p),\\
\end{aligned}
\label{eq:two-loop}
\end{equation}
where $q=(\textbf{q},\omega)$ and $k=(\textbf{k},\Omega)$. Integrating over frequencies, we get 
 \begin{eqnarray}
        \Gamma&=&-\int \frac{dk_+dk_-dq_+dq_-}{(2\pi)^4} \left[\theta(\epsilon_{k+\frac{q}{2}+p})-\theta(-\epsilon_{-k+\frac{q}{2}+p})\right]\frac{\theta(\epsilon_{-k+\frac{q}{2}+p}+\epsilon_{k+q/2+p})-\theta(\epsilon_{q+p})}{(\epsilon_{q+p}-\epsilon_{k+\frac{q}{2}+p}-\epsilon_{-k+\frac{q}{2}+p})^2}\nonumber \\
        &=&-\int\frac{dk_+dk_-dq_+dq_-}{(2\pi)^4} \left[\theta((k_++\frac{q_+}{2}+p_+)(k_-+\frac{q_-}{2}+p_-))-\theta(-(k_+-\frac{q_+}{2}-p_+)(k_--\frac{q_-}{2}-p_-))\right]\nonumber \\
        & \times & \frac{\theta(4k_+k_-+(q_++2p_+)(q_-+2p_-))-\theta((q_++p_+)(q_-+p_-))}{(4k_+k_--q_+q_-+2p_+p_-)^2},
\end{eqnarray}
  where $\theta(x)$ is the step function. 
 Without loss of generality, below we will consider $p_+, p_->0$.   
  This restricts the range of integration 
   over ${\bf k}$ and ${\bf q}$ to
  \begin{equation}
  \begin{aligned}
      \Gamma&=-\left(\int_{|\frac{q_+}{2}+p_+|}^\Lambda \int_{|\frac{q_-}{2}+p_-|}^\Lambda \frac{dk_+dk_-}{4\pi^2}+
      \int^{-|\frac{q_+}{2}+p_+|}_{-\Lambda} \int^{-|\frac{q_-}{2}+p_-|}_{-\Lambda} \frac{dk_+dk_-}{4\pi^2}
      \right)\\
      &\times\left(\int^{\Lambda}_{-p_+}\int_{-\Lambda}^{-p_-}\frac{dq_+dq_-}{4\pi^2}+\int_{-\Lambda}^{-p_+}\int^{\Lambda}_{-p_-}\frac{dq_+dq_-}{4\pi^2}\right)\frac{1}{(4k_+k_--q_+q_-+2p_+p_-)^2}\\
      &-\left(\int_{|\frac{q_+}{2}+p_+|}^\Lambda \int^{-|\frac{q_-}{2}+p_-|}_{-\Lambda} \frac{dk_+dk_-}{4\pi^2}+
      \int^{-|\frac{q_+}{2}+p_+|}_{-\Lambda} \int_{|\frac{q_-}{2}+p_-|}^{\Lambda} \frac{dk_+dk_-}{4\pi^2}
      \right)\\
      &\times\left(\int^{\Lambda}_{-p_+}\int^{\Lambda}_{-p_-}\frac{dq_+dq_-}{4\pi^2}+\int_{-\Lambda}^{-p_+}\int_{-\Lambda}^{-p_-}\frac{dq_+dq_-}{4\pi^2}\right)\frac{1}{(4k_+k_--q_+q_-+2p_+p_-)^2},
  \end{aligned}
  \label{y_888}
  \end{equation}
  where $\Lambda$ is a large momentum cutoff.
  We  did the integration numerically and show the results in  Fig.~\ref{fig:Gamma}. 
  In  Fig.~\ref{fig:Gamma}(a) we show that 
  $\Gamma\approx b'L^2$, where $L=\log{\frac{\Lambda}{|\textbf{p}|}}$.
 To estimate the value of $b'$, we note that the dispersion Eq.~\eqref{eq:VHdisp} breaks rotation symmetry, hence the value of $\Gamma$ depends not only on the magnitude of $\textbf{p}$ but also on its direction 
 relative to $k_\pm$. In Fig.~\ref{fig:Gamma}(b)  we plot $\Gamma$ as a function of $L^2$  for different polar angles of $\textbf{p}$. We find that $b'$  falls in the range between $1.4$ and $1.9$. We cannot get a more precise estimate of $b'$ this way, but we emphasize that it is larger than one,     in agreement with the result  obtained using finite-temperature regularization.
 
\section*{Appendix B.  Fermionic self-energy for an extended VH point}

Equation~\eqref{eq:Z_eps} follows from 
the
 second order perturbation theory. It obeys the correct
power counting $Z_{b\boldsymbol{k}}=b^{-4\epsilon}Z_{\boldsymbol{k}}$ and
recovers the $\epsilon\rightarrow0$ limit of Eq.~\eqref{eq:ZID}.  Let us sketch the main ingredients of the analysis.
As before, we use Eq.~(\ref{eq:extVH}) for the dispersion.
 To compute  the self energy  we need the dynamical polarization $\Pi (\boldsymbol{q}, \omega_m)$. Evaluating it at  finite $\epsilon$, we obtain
 \beq
\Pi (\boldsymbol{q}, \omega_m) = \frac{c}{A |\boldsymbol{q}|^{2\epsilon}} \left (1 + b \frac{|\omega_m|}{A \Lambda |\boldsymbol{q}|^{1+2\epsilon}}\right),
\label{w_1}
\eeq
 where $a, b$ are numbers of order unity.
  The minimum $q_{\text{min}}$, up to which we can treat $\Pi(\boldsymbol{q}, \omega_m)$ as a static, frequency-independent quantity is
  \beq
  q_{\text{min}} \sim \left(\frac{|\omega_m|}{A \Lambda}\right)^{\frac{1}{1+2 \epsilon}}.
  \label{w_2}
  \eeq
  We assume and 
   verify
   that the relevant momenta 
     that contribute to  pairing  
      are indeed 
       larger than 
       $ q_{\text{min}}$. 
The self-energy $\Sigma_{\boldsymbol{k}} ( \omega_m)$ is the convolution of $\Pi$ with the fermionic propagator.
Substituting this form of $\Pi$, we find after straightforward calculation that at $\boldsymbol{k}=\boldsymbol{0}$,
\beq
\Sigma (\omega_m) \sim g \omega_m \left(\frac{A\Lambda^{2+2\epsilon}}{|\omega_m|}\right)^{\frac{4\epsilon}{1+2\epsilon}}
\label{w_3}
\eeq
where $c = O(1)$.
Using (\ref{w_2}) to relate frequencies and momenta and paying attention to the correct symmetry in momentum space, 
 we recover 
 Eq.~\eqref{eq:Z_eps}.

\end{document}